\documentclass[onecolumn,amsmath,amssymb,subfigure,12pt]{revtex4-1}
\usepackage{bm}
\usepackage{graphicx}
\usepackage{hhline}
\usepackage[newitem,newenum]{paralist}
 {\pointedenum\begin{enumerate}}%
 {\end{enumerate}}
 {\pointlessenum\begin{enumerate}}%
 {\end{enumerate}} 
\usepackage{mathrsfs}
\usepackage{amssymb}
\usepackage{bm}
\usepackage[dvips]{color}
\usepackage{color}
\usepackage{fancybox}
\usepackage{fancybox}
\usepackage{fancyhdr}
\begin{document}
\title{Thermal fluctuations of granular gas driven by Gaussian thermostat based on two-point kinetic theory}
\author{Ryosuke Yano}
\affiliation{Department of Advanced Energy, University of Tokyo, 5-1-5 Kashiwanoha, Kashiwa, Chiba, 277-8561, Japan}
\begin{abstract}
In this paper, we investigate thermal fluctuations of the granular gas, which is driven by Gaussian thermostat, on the basis of two-point kinetic theory. In particular, we consider thermal fluctuations of the inelastic variable sphere, which was proposed by Yano [J. Phys. A, 46 (37), 375502 (2013)]. Time correlations of thermal fluctuations of the pressure deviator and two times of the heat flux are calculated on the basis of the two-point kinetic theory and compared with their numerical results, which are calculated using the direct simulation Monte Carlo (DSMC) method. Finally, we compare transport coefficients, which are calculated on the basis of the kinetic theory, with those calculated using the DSMC method.
\end{abstract}

\maketitle

\section{Introduction}
The first study of thermal fluctuations of the granular gas was performed by Goldhirsch and van Noije \cite{Goldhirsch}. In their study \cite{Goldhirsch}, Green-Kubo expression was considered for the viscosity coefficient ($\mu$) of the granular gas on the basis of the scaled Sonine polynomials in Chapman-Enskog method. Afterwards, Dufty and Brey \cite{Dufty} formulated Green-Kubo expression for the thermal conductivity ($\kappa$) and diffusive thermal conductivity ($\eta$) together with $\mu$. Brey \textit{et al}. \cite{Brey} proved that their Green-Kubo expression for the transport coefficients of the granular gas reproduces transport coefficients, which are calculated by Chapman-Enskog method, with some good accuracies under the homogeneous cooling state (HCS), using the direct simulation Monte Carlo (DSMC) method \cite{Bird}, whereas their numerical results also indicated that $\kappa$ and $\eta$, which are calculated on the basis of Green-Kubo expression using the DSMC method, deviate from $\kappa$ and $\eta$, which are calculated by the first order approximations in Chapman-Enskog method, in the range of $0 \le \alpha \le 0.7$ ($\alpha$: restitution coefficient) and $\mu$, which is calculated using the DSMC method, is similar to $\mu$, which is calculated by the first order approximation in Chapman-Enskog method, in all the rage of $\alpha$ ($0 \le \alpha \le 1$). \textcolor{black}{Meanwhile, Garzo, Santos and Montanero \cite{Garzo2} proposed the modified Sonine polynomials in Chapman-Enskog method by expanding the velocity distribution function around not Maxwell-Boltzmann distribution but the zeroth order velocity distribution function, which corresponds to the velocity distribution function under the HCS. On the basis of such modified Chapman-Enskog method, Garzo, Santos and Montanero \cite{Garzo2} obtained $\kappa$ and $\eta$, which are much more similar to $\kappa$ and $\eta$, which were calculated on the basis of Green-Kubo expression using the DSMC method by Brey \textit{et al}. \cite{Brey}, than $\kappa$ and $\eta$, which were calculated using the first order approximation in the conventional Chapman-Enskog method \cite{Brey}.} As a recent study related to thermal fluctuations of the granular gas, the thermally fluctuating hydrodynamics equation, namely, Landau-Lifshitz-Navier-Stokes-Fourier (LLNSF) equation \cite{Landau}, for the granular gas, was formulated using the inelastic Boltzmann equation with the noise term Brey \textit{et al}. \cite{Brey2}.\\
In this paper, we aim to calculate general solutions of time correlations of thermal fluctuations of the pressure deviator and two times of the heat flux for the granular gas driven by Gaussian thermostat, which was proposed by Montanero and Santos \cite{Montanero}. \textcolor{black}{In the previous study on Green-Kubo expression for $\mu$, $\kappa$ and $\eta$ by Dufty and Brey \cite{Dufty}, $\mu$, $\kappa$ and $\eta$ were calculated using ensemble averages of moments, which are obtained using two different fluxes at the same time. Therefore, Green-Kubo expression for $\mu$ and $\kappa$ under the elastic limit were obtained as \cite{Dufty}}
\begin{eqnarray*}
\textcolor{black}{\mu}&\textcolor{black}{=}&\textcolor{black}{\mathcal{P}\int_0^{\hat{\epsilon}} \left<\left<H_{ij}^{(2)}\left(\hat{\epsilon}^\prime\right) ,H_{ij}^{(2)}\left(\hat{\epsilon}^\prime\right)\right>\right> d \hat{\epsilon}^\prime >0, ~~\mathcal{P}\in \mathbb{R}^+},\\
\textcolor{black}{\kappa}&\textcolor{black}{=}&\textcolor{black}{\mathcal{Q}\int_0^{\hat{\epsilon}} \left<\left<H_{i}^{(3)}\left(\hat{\epsilon}^\prime\right), H_{i}^{(3)}\left(\hat{\epsilon}^\prime\right)\right>\right>d \hat{\epsilon}^\prime >0,~~\mathcal{Q}\in \mathbb{R}^+},
\end{eqnarray*}
\textcolor{black}{where $\left<\left<X,Y\right>\right>:=\int_{\mathbb{V}^3} f_{\text{\tiny{MB}}}\left(\hat{\epsilon}^\prime\right) X Y d\bm{c}$ ($\bm{c}\in \mathbb{V}^3 \subseteq \mathbb{R}^3$: velocity of a particle), $H_{ij}^{(2)}$ and $H_i^{(3)}$ are Hermite polynomials, which will be defined in Sec. II, and $\hat{\epsilon}$ and $\hat{\epsilon}^\prime$ are nondimensionalized time \cite{Dufty}. Green-Kubo expression for $\mu$, $\kappa$ and $\eta$ by Dufty and Brey \cite{Dufty} is surely accurate in itself, as confirmed by numerical analysis by Brey \textit{et al}. \cite{Brey}, whereas it is basically different from the conventional Green-Kubo expression for $\mu$ and $\kappa$ such as $\mu \propto \left<p_{ij}(t),p_{ij}(0)\right>$ ($p_{ij}$: pressure deviator) or $\kappa \propto \left<q_{i}(t),q_{i}(0)\right>$ ($q_i$: heat flux) ($t \in \mathbb{R}^+$: time) by Zwanzig \cite{Zwanzig}, because Dufty and Brey \cite{Dufty} calculated the transport coefficients on the basis of one particle distribution function on the basis of Chapman-Enskog method. Consequently, we must consider two particle distribution function to obtain the conventional Green-Kubo expression for the transport coefficients such as $\mu \propto \left<p_{ij}(t),p_{ij}(0)\right>$ or $\kappa \propto \left<q_{i}(t),q_{i}(0)\right>$.}\\
To attain our aim, we extend the two-point kinetic theory for the elastic gas by Tsuge and Sagara \cite{Tsuge} to the granular gas driven by Gaussian thermostat. In particular, we discuss the inelastic variable hard sphere (IVHS), which was proposed by Yano \cite{Yano}, as the component of the granular gas. The IVHS model is useful for understanding of the characteristics of thermal fluctuations of the granular gas driven by Gaussian thermostat, because we can estimate how the characteristics of thermal fluctuations change in accordance with the change of the dependency of the collision frequency on the relative velocity between two colliding granular particles. Additionally, the investigation of thermal fluctuations of the granular gas driven by Gaussian thermostat is interesting, because the investigation of thermal fluctuations under the thermally nonequilibrium steady state is a universal problem in the nonequilibrium statistical mechanics. Time correlations of thermal fluctuations of the pressure deviator and two times of the heat flux for the granular gas driven by Gaussian thermostat are calculated on the basis of the two-point kinetic theory \textcolor{black}{to obtain Green-Kubo expression for the transport coefficients of the IVHS driven by Gaussian thermostat}. These analytical solutions of time correlations of thermal fluctuations of the pressure deviator and two times of the heat flux are compared with DSMC results of time correlations of thermal fluctuations of the pressure deviator and two times of the heat flux. Here, we remind that three cases among the IVHS, namely, inelastic hard sphere (IHS), IVHS with $\Omega=0.6$ and inelastic Maxwell sphere (IMS) \cite{Yano}, are calculated, where $\Omega$ will be defined in Sec. II. Finally, $\mu$, $\kappa$ and $\eta$, which are calculated on the basis of the kinetic theory, are compared with \textcolor{black}{$\mu$, $\kappa$ and $\eta$, which are obtained by our Green-Kubo expression using the DSMC method}. \textcolor{black}{In particular, numerical results of the IMS are interesting, because effects of the nonlinear collisional moments on time correlations of thermal fluctuations of the pressure deviator and two times of the heat flux or cooling rate can be removed in the case of IMS.}\\
This paper is organized as follows. The fourth and sixth order spherically symmetric moments are calculated for the IVHS to define the zeroth order approximation of the velocity distribution function in Sec. II. On the basis of the zeroth order approximation of the velocity distribution function, we discuss the two-point kinetic theory for the granular gas driven by Gaussian thermostat \textcolor{black}{to obtain Green-Kubo expression for the transport coefficients.} in Sec. III. Analytical solutions of time correlations of thermal fluctuations of the pressure deviator and two times of the heat flux are compared with DSMC results in Sec. IV, when the granular gas under Gaussian thermostat is composed of the IHS, IVHS with $\Omega=0.6$ or IMS. Additionally, $\mu$, $\kappa$ and $\eta$, which are calculated on the basis of the kinetic theory, are compared with \textcolor{black}{$\mu$, $\kappa$ and $\eta$, which are obtained by our Green-Kubo expression using the DSMC method}. Finally, we make the concluding remarks in Sec. V.
\section{The fourth and sixth order spherically symmetric moments for granular gas driven by Gaussian thermostat}
Firstly, we calculate the fourth and sixth order spherically symmetric moments, \textcolor{black}{when the granular gas under Gaussian thermostat is composed of the IVHS}. The inelastic Boltzmann equation for the IVHS is written under the spatially homogeneous state as
\begin{eqnarray}
\frac{\partial f(\bm{c},t)}{\partial t}+\frac{\zeta}{2} \frac{\partial C_i f(\bm{c},t)}{\partial c_i}=\mathscr{A} \int_{\mathbb{V}_1^3}\int_0^{2\pi} \int_0^\pi g^{1-\xi} \Biggl(\frac{1}{\alpha^{2-\xi}}f\left(\bm{c}^{\prime\prime}\right)f\left(\bm{c}_1^{\prime\prime}\right)-f\left(\bm{c}\right)f\left(\bm{c}_1\right) \Biggr)\sin \chi d \chi d\epsilon d \bm{c}_1, \nonumber \\
\end{eqnarray}
where $f\left(\bm{c},t\right)$ is the velocity distribution function, $\bm{c}\in \mathbb{V}^3 \subseteq \mathbb{R}^3$ and $\bm{c}_1\in \mathbb{V}_1^3 \subseteq \mathbb{R}^3$ are velocities of two colliding IVHS, $\mathscr{A}(\in \mathbb{R}^+)$ is a constant cross section. $\zeta$ is the cooling rate and $C_i:=c_i-u_i$ ($u_i$: flow velocity) is the peculiar velocity, $g=\left|\bm{c}-\bm{c}_1\right|$ is the magnitude of the relative velocity of two colliding IVHS, $\chi \in \left[0,\pi\right]$ is the deflection angle and $\epsilon \in \left[0,2\pi\right]$ is the scattering angle. We consider $\xi \in \left[0,1\right]$, where $\xi=0$ corresponds to the IHS and $\xi=1$ corresponds to the IMS. Of course, $\left(\bm{c}^{\prime\prime},\bm{c}_1^{\prime\prime} \right) \rightarrow \left(\bm{c},\bm{c}_1\right)$ is obtained after a binary collision. \textcolor{black}{Readers remind that the inelastic Maxwell model (IMM) proposed by Bobylev and Cercignani \cite{Bobylev} is different from the IMS, because the collisional term of the IMM is obtained by $\sqrt{T} \mathscr{B} \int_{\mathbb{V}_1^3} \int_0^{2\pi} \int_0^\pi \left[(1/\alpha)f(\bm{c}^{\prime\prime})f(\bm{c}_1^{\prime\prime})-f(\bm{c})f(\bm{c}_1)\right] d\chi d\epsilon d \bm{c}_1$ ($T$: temperature, $\mathscr{B}\in \mathbb{R}^+$). As a result, the dependency of the collision frequency on the temperature by the IMM is same as that by the IHS and $\sin \chi$ in the right hand side of Eq. (1) is set to unity in the IMM. Finally, the IVHS is a straightforward extension of the VHS, which was proposed by Bird \cite{Bird} to calculate molecules with the inverse power low potential without facing to the angular cut-off problem in the DSMC method, to the inelastic collision, as described in the author's previous study \cite{Yano}. Of course, our assumption of the inverse power law potential for the granular gas is physically unrealistic except for the charged granular particles \cite{Scheffler}, which interact with each other through Coulomb force. Thus, the IVHS is a mathematical model to investigate the characteristics of the IHS, furthermore, as well as the IMM.}\\
The zeroth order approximation of $f\left(\bm{c},t\right)$, namely, $f^{(0)}\left(\bm{c},t\right)$ can be approximated using the fourth and sixth order spherically symmetric moments ($a_4$ and $a_6$) as follows:
\begin{eqnarray}
f^{(0)}\left(\bm{c},t\right)=f_{\mbox{\tiny{MB}}}\left(\bm{c},t\right) \left(1+\frac{1}{120}a_4 H^{(4)}+\frac{1}{5400}a_6 H^{(6)} \right).
\end{eqnarray}
where $f_{\mbox{\tiny{MB}}}\left(\bm{c},t\right)$ is Maxwell-Boltzmann distribution, and $H^{(4)}=v^4-10v^2+15$ and $H^{(6)}=v^6-21v^4+105v^2-105$, in which $\bm{v}:=\bm{C}/(RT)^{1/2}$ ($R$: gas constant). Here, $a_4=(1/\rho)\int_{\mathbb{V}^3} H^{(4)} f\left(\bm{c},t\right) d \bm{c}$ and $a_6=(1/\rho)\int_{\mathbb{V}^3} H^{(6)} f\left(\bm{c},t\right) d \bm{c}$ ($\rho$: density).\\
Substituting $f\left(\bm{c},t\right)=f^{(0)}\left(\bm{c},t\right)$ into both sides of Eq. (1), multiplying both sides of Eq. (1) by $C^2/3$ and integrating over $\mathbb{V}^3$, we obtain
\begin{eqnarray}
\rho R \frac{d T}{d t}+\rho \zeta RT =-\frac{5}{2(5+\Omega)\tau}\left(1-\alpha^2\right) \rho RT,
\end{eqnarray}
where $\Omega:=1-\xi$ and $\tau=p/\mu_{el}\left(\Omega\right)$ ($\mu_{el}\left(\Omega\right)$: viscosity coefficient of the elastic VHS, $p$: static pressure).\\
Gaussian thermostat postulates that the total energy is always conserved. As a result, we obtain the cooling rate $\zeta$ from $d T/dt=0$ in Eq. (3), when the flow velocity is always equal to zero, namely, $\left|\bm{u}\right|=0$, as
\begin{eqnarray}
\zeta=\frac{5}{2\left(5+\Omega\right)\tau}\left(1-\alpha^2\right),
\end{eqnarray}
\textcolor{black}{where $\zeta$ in Eq. (4)  is used throughout the analytical discussion, because contributions of $a_4$ and $a_6$ on the cooling rate are markedly small in cases of the IHS, IVHS with $\Omega=0.6$ and IMS, as described in appendix C.}\\
In later numerical analysis, the definition of $\zeta$ in Eq. (4) is not used, because $|\bm{u}|=0$ is not satisfied owing to velocity fluctuations in the DSMC calculation. Therefor, $\zeta$ is calculated from the energy-loss via inelastic collisions by each time step in the DSMC calculation.\\
Substituting $f\left(\bm{c},t\right)=f^{(0)}\left(\bm{c},t\right)$ into both sides of Eq. (1), multiplying both sides of Eq. (1) by $\textcolor{black}{\rho^{-1}}H^{(4)}$ and $\textcolor{black}{\rho^{-1}}H^{(6)}$, respectively, and integrating over $\mathbb{V}^3$, we obtain
\begin{eqnarray}
\frac{d a_4}{dt}&=& A_0+A_1 a_4+A_2 a_6\nonumber \\
\frac{d a_6}{dt}&=& B_0+B_1 a_4+B_2 a_6.
\end{eqnarray}
where $A_i$ and $B_i$ ($i=0,1,2,$) are functions of $\alpha$ and $\Omega$. In Eq. (5), we neglected nonlinear terms $a_4^2$, $a_4 a_6$ and $a_6^2$ to simplify our discussions. Readers are recommended to read the paper by Santos and Montanero \cite{Santos2} to confirm effects of those nonlinear terms. Concrete forms of $A_i$ and $B_i$ ($i=0,1,2,$) in Eq. (5) are defined by Eqs. (A1)-(A6) in appendix A. Steady solutions of $a_4$ and $a_6$ in Eq. (5) are obtained by $d_t a_4=d_t a_6=0$ in Eq. (5). Finally, steady solutions of $a_4$ and $a_6$ are obtained as
\begin{eqnarray}
&&a_4=\frac{\sum_{i=0}^{11}\beta_i \alpha^i}{\sum_{i=0}^{11}\gamma_i \alpha^i}.\\
&&a_6=\frac{\sum_{i=0}^{8}\beta_i^\prime \alpha^i}{\sum_{i=0}^{8}\gamma_i^\prime \alpha^i}.
\end{eqnarray}
$\beta_i$, $\gamma_i$ in Eq. (6) and $\beta_i^\prime$, $\gamma_i^\prime$ in Eq. (7) are defined by Eqs. (A7)-(A10) in appendix A. $a_4$ and $a_6$ in Eqs. (6) and (7) are equal to those under the HCS. On the basis of Chapman-Enskog method, Brilliantov \textcolor{black}{and P$\ddot{\mbox{o}}$schel} \cite{Brilliantov} obtained $a_4$ and $a_6$ for the IHS under the HCS in a different form from Eqs. (6) and (7).\\
Figures 1, 2 and 3 shows $a_4$ (left frame) and $a_6$ (right frame) versus $\alpha$ for the IHS, IVHS with $\Omega=0.6$ and IMS, respectively. DSMC results of $a_4$ and $a_6$ for the IHS, IVHS with $\Omega=0.6$ and IMS are obtained using 5000 sample particles per a cell in $5 \times 5 $ grids in the square domain $x \in [0,1], y \in [0,1]$, \textcolor{black}{in which the periodic boundary condition is used.}\\
\textcolor{black}{Firstly. we investigate $a_4$ and $a_6$ for the IHS. Of course, $a_4$ and $a_6$ are different from the second and third order cumulants, which are calculated using the Sonine polynomials in Chapman-Enskog method, because we use Grad's method in this paper. Then, $a_4$ is $15$ times of the second order cumulant \cite{Brilliantov} \cite{Santos2} and $a_6$ is $-900/8$ times of the third order cumulant \cite{Brilliantov} \cite{Santos2}. $a_4$ and $a_6$ for the IHS in Eqs. (6) and (7) are, however, different from $15$ times of the second order cumulant and $-900/8$ times of the third order cumulant, which were obtained by Brilliantov and P$\ddot{\mbox{o}}$schel \cite{Brilliantov}. Indeed, calculations of $a_4$ and $a_6$ for the IHS, IVHS with $\Omega=0.6$ and IMS are significant for understanding of changes of $a_4$ and $a_6$ in accordance with the change of $\Omega$.} The left frame of Fig. 1 shows that $a_4$ (symbols), which is calculated using the DSMC method, namely, $\left(a_4\right)_{\text{\tiny{DSMC}}}$ is similar to $a_4$ (dashed-dot line), which is obtained as a steady solution by setting $A_2=0$ in Eq. (5) and equal to $a_4$, \textcolor{black}{which is $15$ times of the second order cumulant obtained by Noije-Ernst \cite{Noije} \cite{Yano}}. $a_4$ in Eq. (6) (solid line) is similar to  $\left(a_4\right)_{\text{\tiny{DSMC}}}$ in the range of $0.6 \le \alpha \le 1$, whereas the difference between $a_4$ in Eq. (6) and  $\left(a_4\right)_{\text{\tiny{DSMC}}}$ increases, as $\alpha$ decreases in the range of $0 \le \alpha \le 0.6$. Additionally, $a_4$ in Eq. (6) is similar to $a_4$ (dashed line), which was obtained by Brilliantov and P$\ddot{\mbox{o}}$schel \cite{Brilliantov}, in the range of $0.4 \le \alpha \le 1$. The right frame of Fig. 1 shows that $a_6$ (symbols), which is calculated using the DSMC method, namely, $\left(a_6\right)_{\text{\tiny{DSMC}}}$ is similar to $a_6$ (solid line) in Eq. (7) in the range of $0.3 \le \alpha \le 1$, whereas $\left(a_6\right)_{\text{\tiny{DSMC}}}$ is similar to $a_6$ (dashed lime), which was obtained by Brilliantov and P$\ddot{\mbox{o}}$schel \cite{Brilliantov}, in the range of $0.5 \le \alpha \le 1$. Therefore, $a_6$ in Eq. (7) is more similar to $\left(a_6\right)_{\text{\tiny{DSMC}}}$ than $a_6$, which was obtained by Brilliantov and P$\ddot{\mbox{o}}$schel \cite{Brilliantov}.\\
Next, we investigate $a_4$ and $a_6$ for the IVHS with $\Omega=0.6$. The left frame of Fig. 2 shows that $\left(a_4\right)_{\text{\tiny{DSMC}}}$ is similar to $a_4$ (dashed line), which is obtained as a steady solution by setting $A_2=0$ in Eq. (5), in the range of $0.4 \le \alpha \le 1$. $a_4$ in Eq. (6) (solid line) is similar to $\left(a_4\right)_{\text{\tiny{DSMC}}}$ in the range of $0.7 \le \alpha \le 1$, whereas the difference between $a_4$ in Eq. (6) and $\left(a_4\right)_{\text{\tiny{DSMC}}}$ increases, as $\alpha$ decreases in the range of $0 \le \alpha \le 0.7$. The right frame of Fig. 2 shows that $\left(a_6\right)_{\text{\tiny{DSMC}}}$ is similar to $a_6$ (solid line) in Eq. (7) in the range of $0.2 \le \alpha \le 1$.\\
Finally, we investigate $a_4$ and $a_6$ for the IMS. The left frame of Fig. 3 shows that $\left(a_4\right)_{\text{\tiny{DSMC}}}$ (symbols) is similar to $a_4$ (solid line) in Eq. (6) in the range of $0.1 \le \alpha \le 1$. Here, we must remind that $A_2=0$ in Eq. (5) is always obtained for the IMS. On the other hand, $a_6$ in Eq. (7) diverges at $\alpha \simeq 0.366$ and becomes negative in the range of $0 \le \alpha < 0.366$, as shown in the right frame of Fig. 3. $a_6$ in Eq. (7) approximates $\left(a_6\right)_{\text{\tiny{DSMC}}}$ with good accuracies in the range of $0.6 \le \alpha \le 1$. To avoid the divergence of $a_6$ in Eq. (7), \textcolor{black}{we should have considered nonlinear collisional moments for the IMS such as $a_4^2$ or $a_4a_6$ in the time evolution of $a_6$ in Eq. (5) \cite{memo}. However, such an inclusion of nonlinear collisional moments such as $a_{4}^2$ or $a_4 a_6$ is beyond the scope of this paper. We conjecture that nonlinear terms $a_6^2$ and $a_{2n}$ $\left(4 \le n\right)$ never appear in the time evolution of $a_6$ in the case of the IMS, as $a_4^2$ and $a_6$ never appear in the time evolution of $a_4$ in the case of the IMS, whereas $a_2^2$ and $a_2 a_4$ are always equal to zero in the time evolution of $a_4$ owing to $a_2=0$.
Of course, we must show the concrete form of the time evolution of $a_6$ using nonlinear terms in our future study. From numerical results of $a_4$ and $a_6$ for the IHS, IVHS with $\Omega=0.6$ and IMS, we conclude that both $a_4$ and $a_6$ at $\alpha=0$ increase in accordance with the decrease of $\Omega$.}\\
On the basis of $a_4$ and $a_6$, we define the zeroth order approximation of the velocity distribution function, namely, $f^{(0)}\left(\bm{c}\right)$, in the next section.
\section{Two-point kinetic theory for granular gas driven by Gaussian thermostat}
The two-point kinetic theory was firstly discussed for the elastic gas by Tsuge and Sagara \cite{Tsuge}. In this section, we consider the two-point kinetic theory for the granular gas driven by Gaussian thermostat. The stochastic inelastic Boltzmann equation is the most fundamental equation in the two-point kinetic theory. In particular, the stochastic inelastic Boltzmann equation under Gaussian thermostat is written as
\begin{eqnarray}
&&B \varrho\left(t,\ell\right)=\left(\frac{\partial}{\partial t}+\bm{c} \frac{\partial }{\partial \bm{x}}+\frac{1}{2}\zeta \frac{\partial}{\partial \bm{c}} \cdot \bm{C}\right)\varrho\left(t,\ell\right)-\mathscr{J}\left(\ell_1|\ell\right)\left[ \varrho\left(t,\ell\right)\varrho\left(t,\hat{\ell}\right)\right]=0,\\
&&\mathscr{J}\left(\ell_1|\ell\right)\left[\varrho\left(t,\ell\right)\varrho\left(t,{\ell}_1\right)\right] \nonumber \\
&=&\mathscr{A}\int_{\mathbb{V}^3} d\bm{c}_1 \int_0^{2\pi} d \epsilon \int_0^\pi d\chi g^{1-\xi} \left\{\frac{1}{\alpha^{2-\xi}}\varrho\left(t,\ell^{\prime\prime}\right)\varrho\left(t,{\ell}_1^{\prime\prime}\right)-\varrho\left(t,\ell\right)\varrho\left(t,{\ell}_1\right) \right\}\sigma \sin \chi, \nonumber \\
\end{eqnarray}
where $\varrho\left(t,\ell\right)$ is the microscopic distribution function defined by
\begin{eqnarray}
\varrho\left(t,\ell\right):=m \sum_{s=1}^N \delta \left[\ell-\ell^{(s)}(t) \right],
\end{eqnarray}
where $s$ is the index of the particles, $N$ is the total number of particles, and $\ell:=\left(\bm{c},\bm{x}\right)$ ($\bm{x} \in \mathbb{R}^3$). \textcolor{black}{Of course, $\left(\ell^{\prime\prime},\ell_1^{\prime\prime}\right) \rightarrow \left(\ell,\ell_1\right)$ is obtained after the binary inelastic collision.} From Eq. (8), we assume $\varrho\left(t,\ell\right) \in C^1(\mathbb{R}^+) \times C^1(\mathbb{V}^3) \times C^1(\mathbb{R}^3)$.\\
For convenience, we set $\xi:=\left(t,\ell\right)$. In $\xi$-space, we consider the correlation between two points $\tilde{\alpha}$ and $\tilde{\beta}$. As a result, definitions $\varrho\left(\tilde{\alpha}\right):=\varrho\left(\xi(\tilde{\alpha})\right)$ and $\varrho(\tilde{\beta}):=\varrho(\xi(\tilde{\beta}))$ are used. At a point $\tilde{\alpha}$, the stochastic inelastic Boltzmann equation under Gaussian thermostat in Eq. (8) is rewritten as
\begin{eqnarray}
B\left(\tilde{\alpha}\right)\varrho\left(\tilde{\alpha}\right)&=&\left[\frac{\partial}{\partial t\left(\tilde{\alpha}\right)}+\bm{c}(\tilde{\alpha})\frac{\partial}{\partial \bm{x}\left(\tilde{\alpha}\right)}+\frac{1}{2}\zeta\left(\tilde{\alpha}\right) \frac{\partial}{\partial \bm{c}\left(\tilde{\alpha}\right)} \cdot \bm{C}\left(\tilde{\alpha}\right)\right]\varrho\left(\tilde{\alpha}\right) \nonumber \\
&&-\mathscr{J}\left({\tilde{\alpha}}_1|\tilde{\alpha}\right)\left[\varrho\left(\tilde{\alpha}\right)\varrho\left({\tilde{\alpha}}_1\right)\right]=0.
\end{eqnarray}
Next, we define the averaged quantity in $\ell$-space using finite volumes $(\bm{x} \in )\Delta V_c (\subseteq \mathbb{R}^3)$ and $(\bm{c} \in )\Delta V( \subseteq \mathbb{V}^3)$ as follows
\begin{eqnarray}
\overline{\psi}&:=&\frac{1}{\Delta V_c\Delta V}\int_{\Delta V_c} d\bm{x}\int_{\Delta V} d\bm{c} \psi, \\
\Delta \psi&:=&\psi-\overline{\psi},
\end{eqnarray}
where $\psi$ is the arbitrary function defined in $\ell$-space.\\
On the basis of Eqs. (12) and (13), we obtain following equation by multiplying $\Delta \varrho(\tilde{\beta})$ by both sides of Eq. (11) and taking average in $\ell$-space,
\begin{eqnarray}
&&\left[\frac{\partial}{\partial t(\tilde{\alpha})}+\bm{c}(\tilde{\alpha})\frac{\partial}{\partial \bm{x}(\tilde{\alpha})}+\frac{1}{2}\zeta(\tilde{\alpha}) \frac{\partial}{\partial \bm{c}(\tilde{\alpha})} \cdot \bm{C}(\tilde{\alpha})\right]\left[\overline{\varrho(\tilde{\alpha})\varrho(\tilde{\beta})}-f(\tilde{\alpha})f(\tilde{\beta})\right]\nonumber \\
&=&\mathscr{J}\left(\tilde{\alpha}|\tilde{\alpha}_1\right)\left[\overline{\varrho(\tilde{\alpha})\varrho(\tilde{\alpha}_1)\varrho(\tilde{\beta})}-\overline{\varrho(\tilde{\alpha})\varrho(\tilde{\alpha}_1)}f(\tilde{\beta}) \right],
\end{eqnarray}
where $f\left(\tilde{\alpha}\right)=\overline{\varrho\left(\tilde{\alpha}\right)} \wedge f(\tilde{\beta})=\overline{\varrho(\tilde{\beta})}$.\\
In Eq. (14), $\overline{\varrho(\tilde{\alpha})\varrho(\tilde{\beta})}$ is decomposed as
\begin{eqnarray}
\overline{\varrho(\tilde{\alpha})\varrho(\tilde{\beta})}=f_{II}(\tilde{\alpha};\tilde{\beta})+g(\tilde{\alpha};\tilde{\beta}),
\end{eqnarray}
where $f_{II}\left(\tilde{\alpha};\tilde{\beta}\right)$ is the two-point phase density of different particles and $g(\tilde{\alpha};\tilde{\beta})$ is the probability of finding same particles at $t=t(\tilde{\alpha})\wedge \ell=\ell(\tilde{\alpha})$ and $t=t(\tilde{\beta}) \wedge \ell=\ell(\tilde{\beta})$. $f_{II}(\tilde{\alpha};\tilde{\beta})$ is related to the hydrodynamic fluctuation term $\phi(\tilde{\alpha};\tilde{\beta})$ as follows
\begin{eqnarray}
f_{II}(\tilde{\alpha};\tilde{\beta})&=&\phi(\tilde{\alpha};\tilde{\beta})+\left(1+\frac{1}{N}\right)f(\tilde{\alpha})f(\tilde{\beta}),\\
&\simeq& \phi(\tilde{\alpha};\tilde{\beta})+f(\tilde{\alpha})f(\tilde{\beta}), ~~(\mbox{for }1 \ll N),
\end{eqnarray}
For $1 \ll N$ and $t(\tilde{\alpha}) \rightarrow t(\tilde{\beta})$, we obtain
\begin{eqnarray}
&&\lim_{t(\tilde{\alpha}) \rightarrow t(\tilde{\beta})} f_{II}\left(\tilde{\alpha};\tilde{\beta}\right)=\phi(\tilde{\alpha},\tilde{\beta})+f\left(\tilde{\alpha}\right)f(\tilde{\beta}),\\
&&\lim_{t\left(\tilde{\alpha}\right) \rightarrow t(\tilde{\beta})} g\left(\tilde{\alpha};\tilde{\beta}\right)=\delta\left[\xi\left(\tilde{\alpha}\right)-\xi(\tilde{\beta})\right]f\left(\tilde{\alpha}\right).
\end{eqnarray}
Assuming that hydrodynamic fluctuations are ignorable, (i.e., $\phi\left(\tilde{\alpha};\tilde{\beta}\right)=0$ or $f_{II}\left(\tilde{\alpha};\tilde{\beta}\right)=f\left(\tilde{\alpha}\right)f(\tilde{\beta})$), we obtain the following equation from Eq. (14) \cite{Tsuge}
\begin{eqnarray}
&&\left[\frac{\partial}{\partial \epsilon}+\bm{c}(\tilde{\alpha})\frac{\partial}{\partial \bm{x}(\tilde{\alpha})}+\frac{1}{2}\zeta(\tilde{\alpha}) \frac{\partial}{\partial \bm{c}(\tilde{\alpha})} \cdot \bm{C}(\tilde{\alpha})\right]g\left(\tilde{\alpha};\tilde{\beta}\right) \nonumber \\
&&=\mathscr{J}\left(\tilde{\alpha}|{\tilde{\alpha}}_1\right)\left(f(\tilde{\alpha})g({\tilde{\alpha}}_1;\tilde{\beta})+f({\tilde{\alpha}}_1)g(\tilde{\alpha};\tilde{\beta})\right), \nonumber \\
\end{eqnarray}
where $\epsilon := t\left(\tilde{\alpha}\right)-t(\tilde{\beta})$.\\
$g(\tilde{\alpha};\tilde{\beta})$ is expanded using Hermite polynomials \cite{Tsuge} as follows
\begin{eqnarray}
g\left(\tilde{\alpha};\tilde{\beta}\right)=\omega\left(\tilde{\alpha}\right)\omega(\tilde{\beta})\sum\frac{Q^{(J,K)}_{ij...,lm...}}{\tilde{c}(\tilde{\alpha})^{J} \tilde{c}(\tilde{\beta})^K J!K!}H_{ij...}^{(J)}(\tilde{\alpha})H_{lm...}^{(K)}(\tilde{\beta})
\end{eqnarray}
where $\tilde{c}(\tilde{\alpha}):=\sqrt{RT(\alpha)}$, $\tilde{c}(\tilde{\beta}):=\sqrt{RT(\beta)}$, $\omega\left(\tilde{\alpha}\right)=\left(2\pi \tilde{c}(\tilde{\alpha})^2\right)^{-3/2}\exp\left[-v\left(\tilde{\alpha}\right)^2\right]$, \\$\omega(\tilde{\beta})=\left(2\pi \tilde{c}(\tilde{\beta})^2\right)^{-3/2}\exp[-v(\tilde{\beta})^2]$ and $H_{ij...}^{(J)}$ is the Hermite polynomial \cite{Grad}.\\
From Eq. (21), we readily obtain
\begin{eqnarray}
Q_{ij...,lm...}^{(J,K)}\left(\ell(\tilde{\alpha})-\ell(\tilde{\beta}),\epsilon\right)=\tilde{c}(\tilde{\alpha})^{J} \tilde{c}(\tilde{\beta})^K\int H_{ij...}^{(J)}(\tilde{\alpha})H_{lm...}^{(K)}(\tilde{\beta})g(\tilde{\alpha},\tilde{\beta})dv(\tilde{\alpha})dv(\tilde{\beta}).
\end{eqnarray}
We assume the initial form of $g(\tilde{\alpha},\tilde{\beta})$ from Eq. (19) as follows
\begin{eqnarray}
\left[g(\tilde{\alpha},\tilde{\beta})\right]_{\epsilon=0}=\delta\left[\xi\left(\tilde{\alpha}\right)-\xi(\tilde{\beta})\right]f^{(0)}\left(\tilde{\alpha}\right),
\end{eqnarray}
where $f^{(0)}\left(\tilde{\alpha}\right)$ is the zeroth order approximation of the velocity distribution function at a point $\tilde{\alpha}$.\\
Substituting $H^{(0)}=1$, $H_i^{(1)}=v_i$, $H_{ij}^{(2)}=v_iv_j-\delta_{ij}v^2/3$, and $H_i^{(3)}=v_i(v^2-5)$ into Eq. (22) with Eq. (23), we obtain 
\begin{eqnarray}
\left[Q^{(0,0)}\right]_{\epsilon=0}&=&\rho(\tilde{\alpha})\delta\left[\bm{x}(\tilde{\alpha})-\bm{x}(\tilde{\beta})\right],\\
\left[{Q}_{i,j}^{(1,1)}\right]_{\epsilon=0}&=&\rho(\tilde{\alpha}) \tilde{c}^2\delta_{ij}\delta\left[\bm{x}(\tilde{\alpha})-\bm{x}(\tilde{\beta})\right],\\
\left[{Q}_{ij,lm}^{(2,2)}\right]_{\epsilon=0}&=&\rho(\tilde{\alpha}) \tilde{c}^4\left(1+\frac{a_4}{15}\right)\left(\delta_{il}\delta_{jm}+\delta_{jl}\delta_{im}-\frac{2}{3}\delta_{ij}\delta_{lm}\right)\delta \left[\bm{x}(\tilde{\alpha})-\bm{x}(\tilde{\beta})\right],\\
\left[{Q}_{i,j}^{(3,3)}\right]_{\epsilon=0}&=&\rho(\tilde{\alpha}) \tilde{c}^6 \delta_{ij}\left(10+\frac{11}{3}a_4+\frac{a_6}{3}\right)\delta\left[\bm{x}(\tilde{\alpha})-\bm{x}(\tilde{\beta})\right],
\end{eqnarray}
where we used the assumption $T(\tilde{\alpha})=T(\tilde{\beta})=T$ and the relation $\tilde{c}=\sqrt{RT}$, which is obtained in the energy conservative system. $Q^{(0,0)}$ corresponds to the two-point-correlation of thermal fluctuations of the density, $Q_{i,j}^{(1,1)}$ corresponds to the two-point-correlation of thermal fluctuations of the momentum, ${Q}_{ij,lm}^{(2,2)}$ corresponds to the two-point-correlation of thermal fluctuations of the pressure deviator, and ${Q}_{i,j}^{(3,3)}$ corresponds to the two-point-correlation of thermal fluctuations of the two times of the heat flux.
Substituting $\hat{H}_{ij}^{(2)}\left(\tilde{\alpha}\right) \hat{H}_{lm}^{(2)}(\tilde{\beta})$ or $H_i^{(3)}\left(\tilde{\alpha}\right)H_j^{(3)}(\tilde{\beta})$ into both sides of Eq. (20) and integrating in accordance with Eq. (22), we obtain time evolutions of $Q_{ij,lm}^{(2,2)}$ and $Q_{i,j}^{(3,3)}$ as follows
\begin{eqnarray}
\frac{d Q_{ij,lm}^{\left(2,2\right)}\left(\epsilon\right)}{d \epsilon}&=&\left(\zeta\left(\tilde{\alpha}\right)-\nu_{\pi}\left(\tilde{\alpha}\right) \right) Q_{ij,lm}^{(2,2)}\left(\epsilon\right), \\
\frac{d Q_{i,j}^{(3,3)}\left(\epsilon\right)}{d \epsilon}&=&\left(\frac{3}{2} \zeta\left(\tilde{\alpha}\right)-\nu_q\left(\tilde{\alpha}\right)\right)Q_{i,j}^{\left(3,3\right)}\left(\epsilon\right),
\end{eqnarray}
where we restrict ourselves to the spatially homogeneous state. \textcolor{black}{Equations (28) and (29) are equivalent to Eqs. (C3) and (C4), when we replace $Q_{ij,lm}^{(2,2)}$ and $Q_{i,j}^{(3,3)}$ in Eqs. (28) and (29) with $p_{ij}$ and $q^i$, respectively, and neglect all the nonlinear terms in Eqs. (C3) and (C4). As a result, Eqs. (28) and (29) includes nonlinear terms except for the IMS, as shown in Eqs. (C3) and (C4). We, however, neglect such nonlinear terms as small amounts in Eqs. (28) and (29), as described in appendix C.}\\
From Eqs. (28) and (29), we obtain
\begin{eqnarray}
Q_{ij,lm}^{(2,2)}(\epsilon)&=&\left[Q_{ij,lm}^{(2,2)}\right]_{\epsilon=0} \exp\left[-\left(\nu_{\pi}\left(\tilde{\alpha}\right)-\zeta\left(\tilde{\alpha}\right)\right) \epsilon\right],\\
Q_{i,j}^{(3,3)}(\epsilon)&=&\left[Q_{i,j}^{(3,3)}\right]_{\epsilon=0} \exp\left[-\left(\nu_q\left(\tilde{\alpha}\right)-\frac{3}{2}\zeta\left(\tilde{\alpha}\right)\right) \epsilon \right].
\end{eqnarray}
where $\nu_{\pi}$ and $\nu_{q}$ are dissipation rates of $Q_{ij,lm}^{(2,2)}$ and $Q_{i,j}^{(3,3)}$, which are calculated as \cite{Yano}
\begin{eqnarray}
\nu_{\pi}&=&\frac{1}{4\tau}\left(3-\alpha\right)\left(1+\alpha\right), \\
\nu_q&=&-\frac{1}{\tau}\frac{\left(1+\alpha\right)\left(110+37\Omega-\alpha(70+29\Omega)\right)}{24\left(5+\Omega\right)}.
\end{eqnarray}
Provided that $ 1 \ll \nu_{\pi}\left(\tilde{\alpha}\right)-\zeta\left(\tilde{\alpha}\right)$ and $1 \ll \nu_q\left(\tilde{\alpha}\right)-\frac{3}{2}\zeta\left(\tilde{\alpha}\right)$, we can approximate $\exp(-A \epsilon)$ with $2/A \delta\left(\epsilon\right)$, so that Eqs. (30) and (31) can be rewritten as
\begin{eqnarray}
Q_{ij,lm}^{\left(2,2\right)}\left(\epsilon\right)&=&\left[Q_{ij,lm}^{\left(2,2\right)}\right]_{\epsilon=0} \frac{2}{\nu_{\pi}\left(\tilde{\alpha}\right)-\zeta\left(\tilde{\alpha}\right)} \delta\left(\epsilon\right),\\
Q_{i,j}^{(3,3)}\left(\epsilon\right)&=&\left[Q_{i,j}^{\left(3,3\right)}\right]_{\epsilon=0}\frac{2}{\nu_q\left(\tilde{\alpha}\right)-\frac{3}{2}\zeta\left(\tilde{\alpha}\right)}\delta\left(\epsilon\right).
\end{eqnarray}
Substituting $\alpha=1$ into Eqs. (34) and (35), respectively, we can readily reproduce thermally fluctuating terms in LLNSF equation for the elastic gas \cite{Landau}.\\
Integrating Eqs. (30) and (31) from $\epsilon=0$ to $\infty$, we obtain
\begin{eqnarray}
\int_0^\infty Q_{ij,lm}^{\left(2,2\right)}\left(\epsilon\right) d\epsilon &=&\left[Q_{ij,lm}^{\left(2,2\right)}\right]_{\epsilon=0} \frac{1}{\nu_{\pi}\left(\tilde{\alpha}\right)-\zeta\left(\tilde{\alpha}\right)},\\
\int_0^\infty Q_{i,j}^{\left(3,3\right)}\left(\epsilon\right) d\epsilon &=&\left[Q_{i,j}^{\left(3,3\right)}\right]_{\epsilon=0}\frac{1}{\nu_q\left(\tilde{\alpha}\right)-\frac{3}{2}\zeta\left(\tilde{\alpha}\right)}.
\end{eqnarray}
Here, we remind that $\mu\left(\alpha,\Omega\right)$, $\kappa\left(\alpha,\Omega\right)$ and $\eta\left(\alpha,\Omega\right)$ for the granular gas driven by Gaussian thermostat are written as \cite{Santos}
\begin{eqnarray}
\mu\left(\alpha,\Omega\right)&=&\frac{p\left(\tilde{\alpha}\right)}{\nu_{\pi}\left(\tilde{\alpha}\right)-\zeta\left(\tilde{\alpha}\right)},\\
\kappa\left(\alpha,\Omega\right)&=&\left(\frac{5}{2}+\frac{a_4}{3}\right)\frac{p\left(\tilde{\alpha}\right)R}{\nu_q\left(\tilde{\alpha}\right)-\frac{3}{2}\zeta\left(\tilde{\alpha}\right)},\\
\eta\left(\alpha,\Omega\right)&=&\frac{a_4}{6}\frac{p\left(\tilde{\alpha}\right)RT\left(\tilde{\alpha}\right)}{\rho\left(\tilde{\alpha}\right)}\frac{1}{\nu_q\left(\tilde{\alpha}\right)-\frac{3}{2}\zeta\left(\tilde{\alpha}\right)},
\end{eqnarray}
\textcolor{black}{where $\mu\left(\alpha,1\right)$, $\kappa\left(\alpha,1\right)$ and $\eta\left(\alpha,1\right)$ for the IHS under Gaussian thermostat are quite same as those obtained by Santos \cite{Santos}.}\\
From Eqs. (38)-(40), we obtain
\begin{eqnarray}
\mu\left(\alpha,\Omega\right)&=&p\left(\tilde{\alpha}\right) \int_0^\infty Q_{ij,lm}^{(2,2)}\left(\epsilon\right) d\epsilon\left[Q_{ij,lm}^{\left(2,2\right)}\right]_{\epsilon=0}^{-1},\\
\kappa\left(\alpha,\Omega\right)&=&\left(\frac{5}{2}+\frac{a_4}{3}\right)  p\left(\tilde{\alpha}\right) R \int_0^\infty Q_{i,j}^{\left(3,3\right)}\left(\epsilon\right) d\epsilon \left[Q_{i,j}^{(3,3)}\right]_{\epsilon=0}^{-1},\\
\eta\left(\alpha,\Omega\right)&=&\frac{a_4}{6} \frac{p\left(\tilde{\alpha}\right)RT\left(\tilde{\alpha}\right)}{\rho\left(\tilde{\alpha}\right)}\int_0^\infty Q_{i,j}^{(3,3)}\left(\epsilon\right) d\epsilon  \left[Q_{i,j}^{(3,3)}\right]_{\epsilon=0}^{-1}.
\end{eqnarray}
Eqs. (41)-(43) are definitions of the transport coefficients of the granular gas driven Gaussian thermostat. Substituting $\alpha=1$ into Eq. (41)-(43), we can readily obtain Green-Kubo expression for the transport coefficients in the elastic gas \textcolor{black}{by Zwanzig \cite{Zwanzig}}.\\
The comparison of the viscosity coefficient of the granular gas driven by Gaussian thermostat, which is calculated using the DSMC method, with that obtained by Chapman-Enskog method was done using the solution of the uniform shear flow by Garzo and Montanero \cite{Garzo}. In Sec. IV, we calculate following four parameters using the DSMC method
\begin{eqnarray*}
\psi_2\left(\alpha\right)&=&\left[Q_{xx,xx}^{(2,2)}\right]_{\epsilon=0}(\alpha)/\left[Q_{xx,xx}^{(2,2)}\right]_{\epsilon=0}^{\alpha=1},\\
\psi_3\left(\alpha\right)&=&\left[Q_{x,x}^{(3,3)}\right]_{\epsilon=0}(\alpha)/\left[Q_{x,x}^{(3,3)}\right]_{\epsilon=0}^{\alpha=1},\\
\phi_2\left(\epsilon\right)&=&Q_{xx,xx}^{(2,2)}\left(\epsilon\right)\left[Q_{xx,xx}^{(2,2)}\right]^{-1}_{\epsilon=0},\\
\phi_3\left(\epsilon\right)&=&Q_{x,x}^{(3,3)}\left(\epsilon\right)\left[Q_{x,x}^{(3,3)}\right]^{-1}_{\epsilon=0}.
\end{eqnarray*}
Analytical solutions of $\psi_2\left(\alpha\right)$ and $\psi_3\left(\alpha\right)$ are readily calculated from Eqs. (26) and (27) as
\begin{eqnarray}
\psi_2\left(\alpha\right)&=&1+\frac{a_4}{15},\\
\psi_3\left(\alpha\right)&=&1+\frac{11}{30}a_4+\frac{a_6}{30}.
\end{eqnarray}
\textcolor{black}{$\psi_2\left(\alpha\right)$ and $\psi_3\left(\alpha\right)$ in Eqs. (44) and (45) hold true for the IVHS under the HCS from our discussion in appendix B.}\\
Similarly, analytical solutions of $\phi_2$ and $\phi_3$ are readily calculated from Eqs. (30) and (31) as
\begin{eqnarray}
\phi_2\left(\epsilon\right)&=&\exp\left[-\left(\nu_{\pi}\left(\tilde{\alpha}\right)-\zeta\left(\tilde{\alpha}\right)\right) \epsilon\right],\\
\phi_3\left(\epsilon\right)&=&\exp\left[-\left(\nu_q\left(\tilde{\alpha}\right)-\frac{3}{2}\zeta\left(\tilde{\alpha}\right)\right) \epsilon \right].
\end{eqnarray}
In Sec. III, we considered time correlations of thermal fluctuations of the pressure deviator and two times of the heat flux for the granular gas driven by Gaussian thermostat on the basis of the two-point kinetic theory, whereas time correlations of thermal fluctuations of the pressure deviator and two times of the heat flux for the granular gas under the HCS are discussed on the basis of the two-point kinetic theory in appendix B.
\section{Comparison of DSMC results with analytical results}
The DSMC calculation is performed to confirm the validity of the two-point kinetic theory for the granular gas driven by Gaussian thermostat. In the DSMC calculation, 5000 sample particles are set in a cell, whereas equally spaced $5 \times 5 $ grids are set in the square domain $x \in [0,1], y \in [0,1]$. All the physical quantities are normalized as: $\hat{\epsilon}=\epsilon/t_\infty$, $\hat{\rho}=\rho/\rho_\infty$, $\hat{T}=T/T_\infty$, $\hat{Q}_{xx,xx}^{(2,2)}=Q_{xx,xx}^{(2,2)}\rho_\infty^{-1} (RT_\infty)^{-2}$ and $\hat{Q}_{x,x}^{(3,3)}=Q_{x,x}^{(3,3)}\rho_\infty^{-1} (RT_\infty)^{-3}$. The time interval is set as $2.5 \times 10^{-5}$ and $2.0\times 10^5$ steps are iterated. $\mbox{Kn}=2.5 \sqrt{2} \times 10^{-4}$ ($\mbox{Kn}$: Knudsen number) for the IHS, $\mbox{Kn}=2.5 (2)^{3/5} \times 10^{-4}$ ($\mbox{Kn}$: Knudsen number) for the IVHS with $\Omega=0.6$ and $\mbox{Kn}=2.5\times 10^{-4}$ for the IMS. DSMC calculations are performed using $\alpha=0, 0.1, 0.2, 0.3, 0.4, 0.5, 0.6, 0.7, 0.8, 0.9, 0.95$ and $1$ for the IHS, IVHS with $\Omega=0.6$ and IMS. \textcolor{black}{Finally, we used the periodic boundary condition, whereas the investigation of effects of the area of the calculation-domain on the numerical results is necessary for the validation of the present numerical results, whereas our numerical results indicate that the long range correlation of thermal fluctuations beyond the identified cell is small in comparison with that inside the identified cell owing to the small (local) Knudsen number. We, however, set such an investigation to our future work, because further increase of grid points requires the parallel computation. In particular, the boundary condition is a significant factor, which characterizes the dynamics of the granular gas \cite{Brey3} \cite{Rouyer} \cite{Zon}.}
\subsection{DSMC results of $\psi_2(\alpha)$ and $\psi_3(\alpha)$ and their comparisons with Eqs. (44) and (45)}
The DSMC results show $\left[\hat{Q}_{xx,xx}^{(2,2)}\right]_{\epsilon=0}=1.33$ and $\left[\hat{Q}_{x,x}^{(3,3)}\right]_{\epsilon=0}=10$ for the IHS, IVHS with $\Omega=0.6$ and IMS, when $\alpha=1$, namely, $a_4=a_6=0$ in Eqs. (26) and (27). Therefore, DSMC results reproduce Eqs. (26) and (27) with good accuracies, when $\alpha=1$.\\
\textcolor{black}{Figures 4, 5 and 6} show $\psi_2\left(\alpha\right)$ versus $\alpha$ (left frame) and $\psi_3\left(\alpha\right)$ versus $\alpha$ (right frame) for the IHS, IVHS with $\Omega=0.6$ and IMS, respectively. $\psi_2\left(\alpha\right)$ in Eq. (44) is calculated using $a_4$ in Eq. (6) or $\left(a_4\right)_{\mbox{\tiny{DSMC}}}$, whereas $\psi_3\left(\alpha\right)$ in Eq. (45) is calculated using $a_4$ in Eq. (6) and $a_6$ in Eq. (7) or $\left(a_4\right)_{\mbox{\tiny{DSMC}}}$ and $\left(a_6\right)_{\mbox{\tiny{DSMC}}}$. The difference between $\left[\psi_2\left(\alpha\right)\right]_{\text{\tiny{DSMC}}}$, which is calculated using the DSMC method, and $\psi_2\left(\alpha\right)$ in Eq. (44), which is obtained using $a_4$ in Eq. (6) or $\left(a_4\right)_{\mbox{\tiny{DSMC}}}$, increases in the range of $0 \le \alpha \le 0.5$, as $\alpha$ decreases. $\psi_2\left(\alpha\right)$ in Eq. (44), which is obtained using $\left(a_4\right)_{\mbox{\tiny{DSMC}}}$, is more similar to $\left[\psi_2\left(\alpha\right)\right]_{\text{\tiny{DSMC}}}$ than $\psi_2\left(\alpha\right)$ in Eq. (44), which is obtained using $a_4$ in Eq. (6). In the right frame of Fig. 4, the difference between $\left[\psi_3\left(\alpha\right)\right]_{\text{\tiny{DSMC}}}$, which is calculated using the DSMC method, and $\psi_3\left(\alpha\right)$ in Eq. (45), which is obtained using $a_4$ in Eq. (6) and $a_6$ in Eq. (7) or $\left(a_4\right)_{\mbox{\tiny{DSMC}}}$ and $\left(a_6\right)_{\mbox{\tiny{DSMC}}}$, increases in the range of $0 \le \alpha \le 0.5$, as $\alpha$ decreases. We, however, find that $\psi_3\left(\alpha\right)$ in Eq. (45), which is obtained using $\left(a_4\right)_{\mbox{\tiny{DSMC}}}$ and $\left(a_6\right)_{\mbox{\tiny{DSMC}}}$, is much more similar to $\left[\psi_3\left(\alpha\right)\right]_{\text{\tiny{DSMC}}}$ than $\psi_3\left(\alpha\right)$ in Eq. (45), which is obtained using $a_4$ in Eq. (6) and $a_6$ in Eq. (7).\\
In the left frame of Fig. 5, the difference between $\left[\psi_2\left(\alpha\right)\right]_{\text{\tiny{DSMC}}}$ and $\psi_2\left(\alpha\right)$ in Eq. (44), which is obtained using $a_4$ in Eq. (6) or $\left(a_4\right)_{\mbox{\tiny{DSMC}}}$, increases in the range of $0 \le \alpha \le 0.6$, as $\alpha$ decreases. $\psi_2\left(\alpha\right)$ in Eq. (44), which is obtained using $\left(a_4\right)_{\mbox{\tiny{DSMC}}}$, is more similar to $\left[\psi_2\left(\alpha\right)\right]_{\text{\tiny{DSMC}}}$ than $\psi_2\left(\alpha\right)$ in Eq. (44), which is obtained using $a_4$ in Eq. (6). The difference between $\left[\psi_3\left(\alpha\right)\right]_{\text{\tiny{DSMC}}}$ and $\psi_3\left(\alpha\right)$ in Eq. (45), which is obtained using $a_4$ in Eq. (6) and $a_6$ in Eq. (7) or $\left(a_4\right)_{\mbox{\tiny{DSMC}}}$ and $\left(a_6\right)_{\mbox{\tiny{DSMC}}}$, increases in the range of $0 \le \alpha \le 0.4$, as $\alpha$ decreases. We, however, find that $\psi_3\left(\alpha\right)$ in Eq. (45), which is obtained using $\left(a_4\right)_{\mbox{\tiny{DSMC}}}$ and $\left(a_6\right)_{\mbox{\tiny{DSMC}}}$, is much more similar to $\left[\psi_3\left(\alpha\right)\right]_{\text{\tiny{DSMC}}}$ than $\psi_3\left(\alpha\right)$ in Eq. (45), which is obtained using $a_4$ in Eq. (6) and $a_6$ in Eq. (7).\\
In the left frame of Fig. 6, the difference between $\left[\psi_2\left(\alpha\right)\right]_{\text{\tiny{DSMC}}}$ and $\psi_2\left(\alpha\right)$ in Eq. (44), which is obtained using $a_4$ in Eq. (6) or $\left(a_4\right)_{\mbox{\tiny{DSMC}}}$, is smaller than that in the case of the IHS. There is only marked difference between $\left[\psi_2\left(\alpha\right)\right]_{\text{\tiny{DSMC}}}$ and $\psi_2\left(\alpha\right)$ in Eq. (44), which is calculated using $\left(a_4\right)_{\mbox{\tiny{DSMC}}}$ or $a_4$ in Eq. (6), when $\alpha=0$. In the right frame of Fig. 6, there are marked differences between $\left[\psi_3\left(\alpha\right)\right]_{\text{\tiny{DSMC}}}$ and $\psi_3\left(\alpha\right)$ in Eq. (45), which is obtained using $\left(a_4\right)_{\mbox{\tiny{DSMC}}}$ and $\left(a_6\right)_{\mbox{\tiny{DSMC}}}$, when $\alpha=0.1$ and $0.2$. Additionally, there are marked differences between $\left[\psi_3\left(\alpha\right)\right]_{\text{\tiny{DSMC}}}$ and $\psi_3\left(\alpha\right)$ in Eq. (45), which is calculated using $a_4$ in Eq. (6) and $a_6$ in Eq. (7), when $0 \le \alpha \le 0.5$, because $a_6$ in Eq. (7) diverges at $\alpha=0.3666$, as shown in the right frame of Fig. 3.\\
\textcolor{black}{In summary, the difference between $a_4$ in Eq. (6) and $\left(a_4\right)_{\text{\tiny{DSMC}}}$ or $a_6$ in Eq. (7) and $\left(a_6\right)_{\text{\tiny{DSMC}}}$ is one of causes of the difference between $\psi_2\left(\alpha\right)$ in Eq. (44) and $\left[\psi_2\left(\alpha\right)\right]_{\text{\tiny{DSMC}}}$ or $\psi_3\left(\alpha\right)$ in Eq. (45) and $\left[\psi_3\left(\alpha\right)\right]_{\text{\tiny{DSMC}}}$.}
\subsection{DSMC results of $\phi_2\left(\hat{\epsilon}\right)$ and $\phi_3\left(\hat{\epsilon}\right)$ and their comparisons with Eqs. (46) and (47)}
We investigate $\phi_2\left(\hat{\epsilon}\right)$ and $\phi_3\left(\hat{\epsilon}\right)$ using the DSMC method and Eqs. (46) and (47). Figures 7, \textcolor{black}{8 and 9} show $\phi_2\left(\hat{\epsilon}\right)$ versus $\hat{\epsilon}$ and $\phi_3\left(\hat{\epsilon}\right)$ versus $\hat{\epsilon}$ for the IHS, IVHS with $\Omega=0.6$ and IMS, when $\alpha=0$, $0.2$, $0.4$, $0.6$, $0.8$ and $1$, respectively. Figure 7 shows that $\left[\phi_2\left(\hat{\epsilon}\right)\right]_{\text{\tiny{DSMC}}}$, which is calculated using the DSMC method, is quite similar to $\phi_2\left(\hat{\epsilon}\right)$ in Eq. (46). Meanwhile, $\left[\phi_3\left(\hat{\epsilon}\right)\right]_{\text{\tiny{DSMC}}}$, which is calculated using the DSMC method, is slightly smaller than $\phi_3\left(\hat{\epsilon}\right)$ in Eq. (47), when $\alpha=1$ and $0.8$, whereas $\left[\phi_3\left(\hat{\epsilon}\right)\right]_{\text{\tiny{DSMC}}}$ is slightly larger than $\phi_3\left(\hat{\epsilon}\right)$ in Eq. (47), when $\alpha=0$. The difference between $\left[\phi_3\left(\hat{\epsilon}\right)\right]_{\text{\tiny{DSMC}}}$ and $\phi_3\left(\hat{\epsilon}\right)$ in Eq. (47), when $\alpha=1$, might be improved by the enhancement of the accuracy of the time integration in the DSMC method. On the other hand, the difference between $\left[\phi_3\left(\hat{\epsilon}\right)\right]_{\text{\tiny{DSMC}}}$ and $\phi_3\left(\hat{\epsilon}\right)$ in Eq. (47), when $\alpha=0$, indicates that the linear response form in Eq. (29) is insufficient to demonstrate $\phi_3\left(\hat{\epsilon}\right)$ for the IHS, when $\alpha=0$.\\
Figure 8 shows that $\left[\phi_2\left(\hat{\epsilon}\right)\right]_{\text{\tiny{DSMC}}}$ is quite similar to $\phi_2\left(\hat{\epsilon}\right)$ in Eq. (46). Meanwhile, $\left[\phi_3\left(\hat{\epsilon}\right)\right]_{\text{\tiny{DSMC}}}$ is slightly smaller than $\phi_3\left(\hat{\epsilon}\right)$ in Eq. (47), when $\alpha=1$, whereas $\left[\phi_3\left(\hat{\epsilon}\right)\right]_{\text{\tiny{DSMC}}}$ is slightly larger than $\phi_3\left(\hat{\epsilon}\right)$ in Eq. (47), when $\alpha=0$. Such tendencies of $\left[\phi_2\left(\hat{\epsilon}\right)\right]_{\text{\tiny{DSMC}}}$ and $\left[\phi_3\left(\hat{\epsilon}\right)\right]_{\text{\tiny{DSMC}}}$ for the IVHS with $\Omega=0.6$ are quite similar to those for the IHS.\\
Figure 9 shows that $\left[\phi_2\left(\hat{\epsilon}\right)\right]_{\text{\tiny{DSMC}}}$ is quite similar to $\phi_2\left(\hat{\epsilon}\right)$ in Eq. (46). Meanwhile, $\left[\phi_3\left(\hat{\epsilon}\right)\right]_{\text{\tiny{DSMC}}}$ is larger than $\phi_3\left(\hat{\epsilon}\right)$ in Eq. (47), when $\alpha=0$ and $0.2$. The difference between $\left[\phi_3\left(\hat{\epsilon}\right)\right]_{\text{\tiny{DSMC}}}$ and $\phi_3\left(\hat{\epsilon}\right)$ in Eq. (47) increases, as $\alpha$ decreases, as shown in frames of $\alpha=0$ and $0.2$. Additionally, we can confirm that the difference between $\left[\phi_3\left(\hat{\epsilon}\right)\right]_{\text{\tiny{DSMC}}}$ and $\phi_3\left(\hat{\epsilon}\right)$ in Eq. (47) when $\alpha=0$ increases, as $\Omega$ decreases, as shown in Figs. 7-9. Finally, we can conclude that the difference between $\left[\phi_3\left(\hat{\epsilon}\right)\right]_{\text{\tiny{DSMC}}}$ and $\phi_3\left(\hat{\epsilon}\right)$ in Eq. (47), when $\alpha=0$ or $0.2$, is not caused by nonlinear terms in the collisional moments of $Q_{x,x}^{(3,3)}$ \cite{Yano} for the IHS and IVHS with $\Omega=0.6$, which are not included in Eq. (29), because nonlinear terms in the collisional moments of $Q_{x,x}^{(3,3)}$ never appear in the right hand side of Eq. (29) for the IMS, \textcolor{black}{as discussed in appendix C}.
\subsection{DSMC results of transport coefficients and their comparisons with Eqs. (38)-(40)}
Finally, we compare transport coefficients, $\mu$, $\kappa$ and $\eta$ in Eqs. (41)-(43), which are calculated using the DSMC method, with those in Eqs. (38)-(40). As shown in Eqs. (41)-(43), we can understand that transport coefficients are related to areas of $\phi_2\left(\hat{\epsilon}\right)$ and $\phi_3\left(\hat{\epsilon}\right)$ in $\hat{\epsilon}\in [0,+\infty)$. Therefore, accuracies of numerical integrations of $\phi_2\left(\hat{\epsilon}\right)$ and $\phi_3\left(\hat{\epsilon}\right)$, which are calculated using the DSMC method, are significant. Meanwhile, $\phi_2\left(\hat{\epsilon}\right)$ and $\phi_3\left(\hat{\epsilon}\right)$ fluctuate around $0$. Therefore, $\phi_2\left(\hat{\epsilon}\right)$ and $\phi_3\left(\hat{\epsilon}\right)$ become negative values in ranges of $|\phi_2\left(\hat{\epsilon}\right)| \ll 1$ and $|\phi_3\left(\hat{\epsilon}\right)| \ll 1$. Such fluctuations around 0 can be reduced by increasing the number of sample particles. We, however, find that further increase of the number of sample particles requires parallel computations. In numerical integrations of $\phi_2\left(\hat{\epsilon}\right)$ and $\phi_3\left(\hat{\epsilon}\right)$, we integrate $\phi_2\left(\hat{\epsilon}\right)$ and $\phi_3\left(\hat{\epsilon}\right)$ in the range of $\left[0,\hat{\epsilon}_c\right]$, in which both $\phi_2\left(\hat{\epsilon}\right)$ and $\phi_3\left(\hat{\epsilon}\right)$ are always positive. In particular, we are interested in $\mu\left(\alpha\right)/\mu(0)$, $\kappa\left(\alpha\right)/\kappa(0)$ and $\tilde{\eta}\left(\alpha\right)=\eta\left(\alpha\right)\rho^{-1}_\infty (RT_\infty)^{-2}$. Additionally, we remind that the transport coefficients do not depend on $\left[Q_{xx,xx}^{(2,2)}\right]_{\hat{\epsilon}=0}$ and $\left[Q_{x,x}^{(3,3)}\right]_{\hat{\epsilon}=0}$, as shown in Eqs. (41)-(43). Consequently, differences between $\left[Q_{xx,xx}^{(2,2)}\right]_{\hat{\epsilon}=0}$ and $\left[Q_{x,x}^{(3,3)}\right]_{\hat{\epsilon}=0}$, which are calculated using the DSMC method, and those in Eqs. (26) and (27) do not contribute to any differences between transport coefficients, which are calculated using the DSMC method in Eqs. (41)-(43), and those in Eqs. (38)-(40).\\
Figure 10 shows $\mu\left(\alpha\right)/\mu(0)$, $\kappa\left(\alpha\right)/\kappa(0)$ and $\tilde{\eta}\left(\alpha\right)$ versus $\alpha$. $a_4$ in Eqs. (39) and (40) are equal to a steady solution of Eq. (5) with $A_2=0$ or Eq. (6). As mentioned above, $a_4$, which is a steady solution of Eq. (5) with $A_2=0$, coincides with $a_4$ in Eq. (6) only for the IMS. Figure 10 shows that $\left[\mu\left(\alpha\right)/\mu(0)\right]_{\text{\tiny{DSMC}}}$, which is calculated using the DSMC method in Eq. (41), is quite similar to that in Eq. (38) for the IHS, IVHS with $\Omega=0.6$ and IMS in all the range of $\alpha$. Such a similarity between $\mu(\alpha)/\mu(0)$ in Eq. (38) and $\left[\mu\left(\alpha\right)/\mu(0)\right]_{\text{\tiny{DSMC}}}$ was also confirmed for the IHS by Garzo and his coworkers \cite{Garzo} \cite{Garzo2}.\\
Figure 10 shows that $\kappa(\alpha)/\kappa(0)$ in Eq. (39) with $a_4$ in Eq. (6) or $a_4$, which is a steady solution of Eq. (5) with $A_2=0$, does not fit $\left[\kappa(\alpha)/\kappa(0)\right]_{\text{\tiny{DSMC}}}$, which is calculated using the DSMC method in Eq. (42), in ranges of $0.4 \le \alpha < 1$ and $0 \le \alpha \le 0.1$ in the case of the IHS. Meanwhile, $\kappa(\alpha)/\kappa(0)$ in Eq. (39) with $a_4$ in Eq. (6) is more similar to $\left[\kappa(\alpha)/\kappa(0)\right]_{\text{\tiny{DSMC}}}$ than $\kappa\left(\alpha\right)/\kappa(0)$ in Eq. (39) with $a_4$, which is a steady solution of Eq. (5) with $A_2=0$, in the range of $0 \le \alpha \le 0.3$ in the case of the IHS. $\kappa(\alpha)/\kappa(0)$ in Eq. (39) with $a_4$ in Eq. (6) or $a_4$, which is a steady solution of Eq. (5) with $A_2=0$, does not fit $\left[\kappa(\alpha)/\kappa(0)\right]_{\text{\tiny{DSMC}}}$ in the range of $0\le \alpha \le 0.7$ in the case of the IVHS with $\Omega=0.6$. Meanwhile, $\kappa(\alpha)/\kappa(0)$ in Eq. (39) with $a_4$ in Eq. (6) is more similar to $\left[\kappa(\alpha)/\kappa(0)\right]_{\text{\tiny{DSMC}}}$ than $\kappa\left(\alpha\right)/\kappa(0)$ in Eq. (39) with $a_4$, which is a steady solution of Eq. (5) with $A_2=0$, in the range of $0 \le \alpha \le 0.7$ in the case of the IVHS with $\Omega=0.6$. $\kappa(\alpha)/\kappa(0)$ in Eq. (39) with $a_4$ in Eq. (6) does not fit $\left[\kappa(\alpha)/\kappa(0)\right]_{\text{\tiny{DSMC}}}$ in the range of $0 \le \alpha \le 0.7$ in the case of the IMS. Meanwhile, the difference between $\left[\kappa(\alpha)/\kappa(0)\right]_{\text{\tiny{DSMC}}}$ and $\kappa(\alpha)/\kappa(0)$ in Eq. (39) with $a_4$ in Eq. (6) increases markedly, as $\alpha$ decreases from $\alpha=0.7$ to $0$, in the case of the IMS. As a result, the difference in the range of $0 \le \alpha \le 0.7$ between $\left[\kappa(\alpha)/\kappa(0)\right]_{\text{\tiny{DSMC}}}$ and $\kappa(\alpha)/\kappa(0)$ in Eq. (39) with $a_4$ in Eq. (6) increases, as $\Omega$ decreases from unity (IHS) to zero (IMS).\\
Figure 10 shows that $\tilde{\eta}(\alpha)$ in Eq. (40) with $a_4$ in Eq. (6) or $a_4$, which is a steady solution of Eq. (5) with $A_2=0$, does not fit $\tilde{\eta}(\alpha)_{\text{\tiny{DSMC}}}$, which is calculated using the DSMC method in Eq. (43), in the range of $0 \le \alpha \le 0.6$ in the case of the IHS. Meanwhile, $\tilde{\eta}(\alpha)$ in Eq. (40) with $a_4$ in Eq. (6) is more similar to $\tilde{\eta}(\alpha)_{\text{\tiny{DSMC}}}$ than $\tilde{\eta}(\alpha)$ in Eq. (40) with $a_4$, which is a steady solution of Eq. (5) with $A_2=0$, in the range of $0 \le \alpha \le 0.5$ in the case of the IHS. $\tilde{\eta}(\alpha)$ in Eq. (40) with $a_4$ in Eq. (6) or $a_4$, which is a steady solution of Eq. (5) with $A_2=0$, does not fit $\tilde{\eta}(\alpha)_{\text{\tiny{DSMC}}}$ in the range of $0 \le \alpha \le 0.6$ in the case of the IVHS with $\Omega=0.6$. Meanwhile, $\tilde{\eta}(\alpha)$ in Eq. (40) with $a_4$ in Eq. (6) is more similar to $\tilde{\eta}(\alpha)_{\text{\tiny{DSMC}}}$ than $\tilde{\eta}(\alpha)$ in Eq. (40) with $a_4$, which is a steady solution of Eq. (5) with $A_2=0$, in the range of $0 \le \alpha \le 0.5$ in the case of the IVHS with $\Omega=0.6$. $\tilde{\eta}(\alpha)$ in Eq. (40) with $a_4$ in Eq. (6) does not fit $\tilde{\eta}(\alpha)_{\text{\tiny{DSMC}}}$ in the range of $0\le \alpha \le 0.6$ in the case of the IMS. Meanwhile, the difference between $\tilde{\eta}(\alpha)_{\text{\tiny{DSMC}}}$ and $\tilde{\eta}(\alpha)$ in Eq. (40) with $a_4$ in Eq. (6) increases markedly, as $\alpha$ decreases from $\alpha=0.5$ to $0$, in the case of the IMS. As a result, the difference in the range of $0 \le \alpha \le 0.5$ between $\tilde{\eta}(\alpha)_{\text{\tiny{DSMC}}}$ and $\tilde{\eta}(\alpha)$ in Eq. (40) with $a_4$ in Eq. (6) increases, as $\Omega$ decreases from unity (IHS) to zero (IMS).
\section{Concluding Remarks}
In this paper, we investigated thermal fluctuations of the granular gas driven by Gaussian thermostat on the basis of the two-point kinetic theory. In particular, we considered the inelastic variable hard sphere (IVHS) as the component of the granular gas. \textcolor{black}{Green-Kubo expression for the transport coefficients, which was proposed in this paper, approximates to that for the elastic gas by Zwanzig under the elastic limit. Therefore, Green-Kubo expression for the transport coefficients, which was proposed in this paper, is different from Green-Kubo expression for the transport coefficients by Dufty and Brey.} Spherically symmetric moment $a_4$, which is calculated using the DSMC method, is more similar to $a_4$, which is analytically obtained by neglecting $a_6$ in the collisional term of $a_4$, than $a_4$, which is analytically obtained by including $a_6$ in the collisional term of $a_4$ in cases of the IHS and IVHS with $\Omega=0.6$. $a_6$, which is calculated using the DSMC method, is similar to $a_6$, which is analytically obtained, in the range of $0.5 \le \alpha \le 1$ in the case of the IHS and in the range of $0.2 \le \alpha \le 1$ in the case of IVHS with $\Omega=0.6$, whereas $a_6$, which is analytically obtained for the IMS, diverges at $\alpha\simeq 0.366$. Correlations of thermal fluctuations of the pressure deviator and two times of the heat flux at the same time were evaluated using two parameters $\psi_2(\alpha)$ and $\psi_3(\alpha)$. $\psi_2(\alpha)$, which is calculated using the DSMC method, is similar to $\psi_2(\alpha)$ in Eq. (44) in the range of $0.6 \le \alpha \le 1$ in cases of the IHS and IVHS, whereas $\psi_2(\alpha)$, which is calculated using the DSMC method, is similar to $\psi_2(\alpha)$ in Eq. (44) in the range of $0.1 \le \alpha \le 1$ in the case of the IMS. Meanwhile, $\psi_3(\alpha)$ in Eq. (45) is similar to $\psi_3(\alpha)$, which is calculated using the DSMC method, in the range of $0.6 \le \alpha \le 1$ in the case of the IHS and in the range of $0.5 \le \alpha \le 1$ in cases of the IVHS with $\Omega=0.6$ and IMS. The use of $(a_4)_{\mbox{\tiny{DSMC}}}$ and $(a_6)_{\mbox{\tiny{DSMC}}}$ in Eqs. (44) and (45) improved similarities between $\psi_2(\alpha)$, which is calculated using the DSMC method, and $\psi_2(\alpha)$ in Eq. (44) with $a_4$ in Eq. (6) or $\psi_3(\alpha)$, which is calculated using the DSMC method, and $\psi_3(\alpha)$ in Eq. (45) with $a_4$ and $a_6$ in Eqs. (6) and (7). Time correlations of thermal fluctuations of the pressure deviator and two times of the heat flux were evaluated using two parameters, namely, $\phi_2\left(\hat{\epsilon}\right)$ and $\phi_3\left(\hat{\epsilon}\right)$. $\phi_2\left(\hat{\epsilon}\right)$, which is calculated using the DSMC method, is similar to $\phi_2\left(\hat{\epsilon}\right)$ in Eq. (46) in cases of the IHS, IVHS with $\Omega=0.6$ and IMS. $\phi_3\left(\hat{\epsilon}\right)$, which is calculated using the DSMC method, is slightly smaller than $\phi_3\left(\hat{\epsilon}\right)$ in Eq. (47), when $\alpha=0$ for the IHS and IVHS with $\Omega=0.6$ or $\alpha=0$ and $0.2$ for the IMS. The viscosity coefficient, which is calculated using the DSMC method, is quite similar to the viscosity coefficient, which is analytically obtained by the kinetic theory, in cases of the IHS, IVHS with $\Omega=0.6$ and IMS. The thermal conductivity, which is calculated using the DSMC method, does not fit the thermal conductivity, which is obtained by the kinetic theory, in ranges of $0.4 \le \alpha \le 1$ and $0 \le \alpha \le 0.1$ in the case of the IHS. The thermal conductivity, which is calculated using the DSMC method, does not fit the thermal conductivity, which is obtained by the kinetic theory, in the range of $0 \le \alpha \le 0.7$ in cases of the IVHS with $\Omega=0.6$ and IMS. The diffusive thermal conductivity, which is calculated using the DSMC method, is similar to the diffusive thermal conductivity, which is obtained by the kinetic theory, in the range of $0.7 \le \alpha \le 1$ in cases of the IHS, IVHS with $\Omega=0.6$ and IMS. Finally, differences between the thermal conductivity and diffusive thermal conductivity at $\alpha=0$, which are calculated using the DSMC method and those obtained by the kinetic theory, increases, as $\Omega$ decreases.
\begin{appendix}
\section{Definitions of symbols in Eqs. (5)-(7)}
$A_i$ and $B_i$ (i=0,1,2) in Eq. (5) are calculated as
\begin{eqnarray}
A_0&=&-\frac{5 \left(\alpha ^2-1\right) \left(\alpha ^2 \left(\Omega +5\right)+2 \Omega -5\right)}{2 \left(\Omega +5\right)},\\
A_1&=&\left(-\alpha^4 \Omega ^3-11 \alpha ^4 \Omega ^2-38 \alpha ^4 \Omega -40 \alpha^4-\alpha^2 \Omega^3+4 \alpha^2 \Omega ^2 \right. \nonumber \\
&& \left.-124 \alpha^2 \Omega +160 \alpha^2+32 \alpha  \Omega +160 \alpha +2 \Omega ^3+7 \Omega ^2+194 \Omega +40\right)\left\{96 \left(\Omega +5\right)\right\}^{-1},\nonumber \\ \\
A_2&=&\left\{\Omega\left(-\alpha ^4 \left(\Omega +2\right)\left(\Omega +4\right)\left(\Omega +5\right)-\alpha ^2 \left(\Omega^3 -4 \Omega^2 +296\Omega+760\right) \right. \right. \nonumber \\
&&\left. \left.+96 \alpha \left(\Omega +5\right)+2 \Omega^3 +7\Omega^2+430\Omega+1280\right)\right\}\left\{8640 \left(\Omega +5\right)\right\}^{-1},\nonumber \\
\\
B_0&=&-\frac{15 \left(\alpha ^2-1\right) \left(\alpha ^4 \left(\Omega +5\right)\left(\Omega +7\right)+2 \alpha ^2 \left(\Omega -7\right)\left(\Omega+5\right)+\Omega \left(3 \Omega -20\right)+35\right)}{16 \left(\Omega +5\right)},\\
B_1&=&\left\{-\alpha ^6 \left(\Omega +4\right)\left(\Omega +5\right)\left(\Omega +6\right)\left(\Omega +7\right)-\alpha ^4 \left(\Omega +5\right)\left(\Omega^3-11\Omega^2+166\Omega+1064\right) \right. \nonumber \\
&&\left. +64 \alpha ^3 \left(\Omega +5\right)\left(\Omega +7\right) +\alpha ^2 \left(15400- \Omega^4-6\Omega^3+289\Omega^2-1318 \Omega\right) \right. \nonumber \\
&&\left.+64 \alpha \left(\Omega -7\right)\left(\Omega +5\right)+3 \Omega^4 +10\Omega^3+707\Omega^2+1854\Omega-9240\right\}\left\{256 \left(\Omega +5\right)\right\}^{-1},\\
B_2&=&-\left\{\alpha ^6 \left(\Omega +2\right)\left(\Omega +4\right)\left(\Omega +5\right)\left(\Omega +6\right)\left(\Omega +7\right) \right. \nonumber \\
&&\left. +\alpha ^4\left(\Omega +2\right)\left(\Omega +5\right)\left(\Omega^3 -11\Omega^2+534\Omega+2856\right) \right. \nonumber \\
&&\left.-192 \alpha ^3 \left(\Omega +2\right)\left(\Omega +5\right)\left(\Omega +7\right) \right. \nonumber \\
&&\left. +\alpha ^2 \left(\Omega^5 -4\Omega^4+581\Omega^3-2084\Omega^2+14796\Omega-114000\right) \right. \nonumber \\
&& \left. -192 \alpha  \left(\Omega ^3+17 \Omega +210\right)-\left(3\Omega^5 +16\Omega^4+1655\Omega^3+8084\Omega^2+56836\Omega+30000\right)\right\} \nonumber \\
&&\left\{23040 \left(\Omega +5\right)\right\}^{-1},
\end{eqnarray}
$\beta_i$ in Eq. (6) are calculated as
\begin{eqnarray}
\beta_0&=&2880 \left(-6 \Omega ^5-335 \Omega ^4-2068 \Omega ^3-13967 \Omega ^2+64830 \Omega -25000\right), \nonumber\\
\beta_2&=&-46080 (\Omega +5) \left(\Omega ^3-10 \Omega ^2+183 \Omega -420\right), \nonumber\\
\beta_3&=&2880 \left(5 \Omega ^5+191 \Omega ^4-1149 \Omega ^3-637 \Omega ^2-176890 \Omega +145000\right),\nonumber\\
\beta_4&=&-46080 \left(\Omega +5\right) \left(3 \Omega ^3+40 \Omega ^2-113 \Omega +700\right),\nonumber\\
\beta_5&=&5760 \left(2 \Omega ^5+208 \Omega ^4+3090 \Omega ^3+13657 \Omega ^2+54950 \Omega -119400\right), \nonumber\\
\beta_6&=&46080 \left(\Omega +5\right) \left(3 \Omega ^3+14 \Omega ^2-13 \Omega +40\right),\nonumber\\
\beta_7&=&-5760 \left(\Omega +5\right) \left(92 \Omega ^3+491 \Omega ^2+538 \Omega -14120\right), \nonumber\\
\beta_8&=&46080 \left(\Omega +4\right)\left(\Omega +5\right)^2 \left(\Omega +7\right), \nonumber\\
\beta_9&=&-2880 \left(\Omega +5\right) \left(2 \Omega ^4+55 \Omega ^3+581 \Omega ^2+2938 \Omega +4200\right), \nonumber\\
\beta_{10}&=&0,\nonumber\\
\beta_{11}&=&-2880 \left(\Omega +2\right)\left(\Omega +4\right)\left(\Omega +5\right)^2 \left(\Omega +7\right),
\end{eqnarray}
$\gamma_i$ in Eq. (6) are calculated as
\begin{eqnarray}
\gamma_0&=&24 \Omega ^7+2460 \Omega ^6+28360 \Omega ^5+361324 \Omega ^4+669696 \Omega ^3+25479248 \Omega ^2+16561280 \Omega -2400000,\nonumber \\
\gamma_1&=&-46080 \left(\Omega+5\right)\left(\Omega^3 -10\Omega^3 +183 \Omega-420\right),\nonumber\\
\gamma_2&=&128 \left(\Omega+5\right)\left(\Omega^5-9\Omega^4+115\Omega^3+1083\Omega^2+75082\Omega-9960\right),\nonumber\\
\gamma_3&=&-20 \Omega^7-1540 \Omega^6+444 \Omega^5-155876 \Omega^4+1711888 \Omega^3-34234800 \Omega^2+20982400 \Omega+12422400,\nonumber\\
\gamma_4&=&384 \left(\Omega +5\right)\left(\Omega^5 +21\Omega^4+95\Omega^3+653\Omega^2-17858\Omega+26280\right),\nonumber\\
\gamma_5&=&-16 \Omega ^7-2784 \Omega ^6-61392 \Omega ^5-536904 \Omega ^4-3784320 \Omega ^3+10602528 \Omega ^2-9758720 \Omega +40896000,\nonumber\\
\gamma_6&=&-384\left(\Omega+5\right) \left(\Omega ^5+11 \Omega ^4+73 \Omega ^3+477 \Omega ^2+1574 \Omega +4200\right),\nonumber\\
\gamma_7&=&8 \left(\Omega +5\right)\left(156 \Omega ^5+1987 \Omega ^4+20240 \Omega ^3+29444 \Omega ^2-258752 \Omega -459840\right),\nonumber\\
\gamma_8&=&-128 \left(\Omega+4\right)\left(\Omega+5\right)^2 \left(\Omega+7\right)\left(\Omega^2 +5\Omega+18\right),\nonumber\\
\gamma_9&=&4 \left(\Omega +4\right)\left(\Omega +5\right) \left(2 \Omega ^5+103 \Omega ^4+1187 \Omega ^3+6288 \Omega ^2+18100 \Omega +21840\right),\nonumber\\
\gamma_{10}&=&0,\nonumber\\
\gamma_{11}&=&4 \left(\Omega +2\right)\left(\Omega +4\right)^2 \left(\Omega +5\right)^2 \left(\Omega +6\right)\left(\Omega +7\right),
\end{eqnarray}
$\beta_i^\prime$ in Eq. (7) are calculated as
\begin{eqnarray}
\beta^\prime_0&=&43200 \left(6 \Omega ^4+213 \Omega ^3+922 \Omega ^2-8435 \Omega +11200\right),\nonumber\\
\beta^\prime_1&=&-86400 \left(6 \Omega ^4+209 \Omega ^3+974 \Omega ^2-8215 \Omega +10500\right),\nonumber\\
\beta^\prime_2&=&43200 \left(13 \Omega ^4+514 \Omega ^3+3941 \Omega ^2-4610 \Omega -11900\right),\nonumber\\
\beta^\prime_3&=&-86400 \left(\Omega +5\right) \left(7 \Omega ^3+258 \Omega ^2+1473 \Omega -4060\right),\nonumber\\
\beta^\prime_4&=&43200 \left(\Omega +5\right) \left(11 \Omega ^3+315 \Omega ^2+1956 \Omega -1820\right),\nonumber\\
\beta^\prime_5&=&-86400 \left(\Omega +5\right) \left(4 \Omega ^3+69 \Omega ^2+507 \Omega +1820\right),\nonumber\\
\beta^\prime_6&=&43200 \left(\Omega +5\right) \left(5 \Omega ^3+81 \Omega ^2+542 \Omega +1820\right),\nonumber\\
\beta^\prime_7&=&-86400 \left(\Omega +4\right)\left(\Omega +5\right)^2 \left(\Omega +7\right),\nonumber\\
\beta^\prime_8&=&43200 \left(\Omega +4\right)\left(\Omega +5\right)^2\left(\Omega +7\right),
\end{eqnarray}
$\gamma^\prime_i$ in Eq. (7) are calculated as
\begin{eqnarray}
\gamma^\prime_0&=&6 \Omega ^7+615 \Omega ^6+7090 \Omega ^5+90331 \Omega ^4+167424 \Omega ^3+6369812 \Omega ^2+4140320 \Omega -600000, \nonumber\\
\gamma^\prime_1&=&-12 \Omega ^7-1198 \Omega ^6-14308 \Omega ^5-178422 \Omega ^4-281792 \Omega ^3-10163720 \Omega ^2+3413760 \Omega -393600,\nonumber\\
\gamma^\prime_2&=&13 \Omega ^7+1396 \Omega ^6+21637 \Omega ^5+227544 \Omega ^4+824132 \Omega ^3+5398928 \Omega ^2-5722240 \Omega +4492800,\nonumber\\
\gamma^\prime_3&=&-2 \left(\Omega +5\right) \left(7 \Omega ^6+714 \Omega ^5+9665 \Omega ^4+80408 \Omega ^3+227052 \Omega ^2-117728 \Omega -402240\right),\nonumber\\
\gamma^\prime_4&=&\left(\Omega +5\right) \left(11 \Omega ^6+849 \Omega ^5+11710 \Omega ^4+94612 \Omega ^3+273096 \Omega ^2-43648 \Omega -462720\right),\nonumber\\
\gamma^\prime_5&=&-2 \left(\Omega +4\right)\left(\Omega +5\right) \left(4 \Omega ^5+167 \Omega ^4+1905 \Omega ^3+10088 \Omega ^2+28588 \Omega +35280\right),\nonumber\\
\gamma^\prime_6&=&\left(\Omega +4\right)\left(\Omega +5\right) \left(5 \Omega ^5+175 \Omega ^4+1856 \Omega ^3+9276 \Omega ^2+24448 \Omega +26880\right),\nonumber\\
\gamma^\prime_7&=&-2 \left(\Omega +2\right)\left(\Omega +4\right)^2 \left(\Omega +5\right)^2 \left(\Omega +6\right)\left(\Omega +7\right),\nonumber\\
\gamma^\prime_8&=&-2 \left(\Omega +2\right)\left(\Omega +4\right)^2\left(\Omega +5\right)^2\left(\Omega +6\right)\left(\Omega +7\right),
\end{eqnarray}
\section{Two-point kinetic theory for granular gas under HCS}
In this paper, we discussed the extension of the two-point kinetic theory by Tsuge-Sagara to the granular gas driven by Gaussian thermostat. In the two-point kinetic theory by Tsuge-Sagara, the form of the correlation function, namely, $g\left(\tilde{\alpha};\tilde{\beta}\right)$, is expanded using Hermite polynomials, as shown in Eq. (21). The validity of the application of the two-point kinetic theory to thermal fluctuations of the granular gas under the HCS requires further considerations. For convenience, we expand $g\left(\tilde{\alpha};\tilde{\beta}\right)$ using modified moments $\tilde{Q}_{ij...,lm...}^{(J,K)}$ such as
\begin{eqnarray}
g\left(\tilde{\alpha};\tilde{\beta}\right)=\omega\left(\tilde{\alpha}\right)\omega(\tilde{\beta})\sum\frac{\tilde{Q}^{(J,K)}_{ij...,lm...}}{J!K!}H_{ij...}^{(J)}(\tilde{\alpha})H_{lm...}^{(K)}(\tilde{\beta})
\end{eqnarray}
From Eq. (B1), we readily obtain
\begin{eqnarray}
\tilde{Q}_{ij...,lm...}^{(J,K)}\left(\ell(\tilde{\alpha})-\ell(\tilde{\beta}),\epsilon\right)=\int H_{ij...}^{(J)}(\tilde{\alpha})H_{lm...}^{(K)}(\tilde{\beta})g(\tilde{\alpha},\tilde{\beta})dv(\tilde{\alpha})dv(\tilde{\beta}).
\end{eqnarray}
From Eqs. (B1), (B2) and (23), we obtain following relations using similar procedures to obtain Eqs. (24)-(27).
\begin{eqnarray}
\left[\tilde{Q}^{(0,0)}\right]_{\epsilon=0}&=&\rho(\tilde{\alpha})\delta\left[\bm{x}(\tilde{\alpha})-\bm{x}(\tilde{\beta})\right],\\
\left[\tilde{Q}_{i,j}^{(1,1)}\right]_{\epsilon=0}&=&\rho(\tilde{\alpha})\delta_{ij}\delta\left[\bm{x}(\tilde{\alpha})-\bm{x}(\tilde{\beta})\right],\\
\left[\tilde{Q}_{ij,lm}^{(2,2)}\right]_{\epsilon=0}&=&\rho(\tilde{\alpha}) \left(1+\frac{a_4}{15}\right)\left(\delta_{il}\delta_{jm}+\delta_{jl}\delta_{im}-\frac{2}{3}\delta_{ij}\delta_{lm}\right)\delta \left[\bm{x}(\tilde{\alpha})-\bm{x}(\tilde{\beta})\right],\\
\left[\tilde{Q}_{i,j}^{(3,3)}\right]_{\epsilon=0}&=&\rho(\tilde{\alpha}) \delta_{ij}\left(10+\frac{11}{3}a_4+\frac{a_6}{3}\right)\delta\left[\bm{x}(\tilde{\alpha})-\bm{x}(\tilde{\beta})\right].
\end{eqnarray}
In Eq. (B2), $\tilde{Q}_{ij...,lm...}^{(J,K)}$ depends on the only density. As a result, Eqs. (B3)-(B6) hold true for the granular gas under the HCS. On the other hand, the calculation of $\left[\tilde{Q}_{ij...,...}^{(J,K)}\right]_{\epsilon=0}$ under the HCS is difficult, owing to the emergence of the long wave instability via the inelastic clustering. Therefore, the choice of the time interval to sample $\left[\tilde{Q}_{ij...,....}^{(J,K)}\right]_{\epsilon=0}$ is significant.\\ 
Finally, we must investigate whether the extension of Green-Kubo expression for the transport coefficients in Eqs. (41)-(43) can be possible for the granular gas under the HCS. According to the two-point kinetic theory in Sec. III, time evolutions of $\tilde{Q}_{ij,lm}^{(2,2)}(\epsilon)$ and $\tilde{Q}_{i,j}^{(3,3)}(\epsilon)$ are calculated as
\begin{eqnarray}
\frac{d \tilde{Q}_{ij,lm}^{\left(2,2\right)}\left(\epsilon\right)}{d \epsilon}&=&-\nu_{\pi}\left(\tilde{\alpha}\right)  \tilde{Q}_{ij,lm}^{\left(2,2\right)}\left(\epsilon\right), \\
\frac{d \tilde{Q}_{i,j}^{(3,3)}\left(\epsilon\right)}{d \epsilon}&=&-\nu_q\left(\tilde{\alpha}\right) \tilde{Q}_{i,j}^{\left(3,3\right)}\left(\epsilon\right).
\end{eqnarray}
The cooling rate $\zeta$ never emerges in Eqs. (B7) and (B8) unlike Eqs. (28) and (29), whereas $\nu_\pi$ and $\nu_q$ are function of time, because the temperature depends on $\epsilon$.\\
Solutions of Eqs. (B7) and (B8) are obtained as
\begin{eqnarray}
\tilde{Q}_{ij,lm}^{\left(2,2\right)}\left(\epsilon\right)&=&\exp\left(-\int_0^\epsilon \nu_{\pi}\left(\tilde{\alpha}\right) ds\right)\left[\tilde{Q}_{ij,lm}^{\left(2,2\right)}\right]_{\epsilon=0},\\
\tilde{Q}_{i,j}^{(3,3)}\left(\epsilon\right)&=&\exp\left(-\int_0^\epsilon \nu_{q}\left(\tilde{\alpha}\right) ds\right)\left[\tilde{Q}_{i,j}^{(3,3)}\right]_{\epsilon=0},
\end{eqnarray}
where $s \in t$.\\
From Eqs. (B9) and (B10), we obtain
\begin{eqnarray}
\int_0^t \tilde{Q}_{ij,lm}^{(2,2)}\left(\epsilon\right) d\epsilon&=&-\left[\nu_{\pi}^{-1}\left(\tilde{\alpha},\epsilon\right)\exp\left(-\int_0^\epsilon \nu_{\pi}\left(\tilde{\alpha},s\right) ds\right)\right]_0^t \left[\tilde{Q}_{ij,lm}^{(2,2)}\right]_{\epsilon=0},\\
\int_0^t \tilde{Q}_{i,j}^{(3,3)}\left(\epsilon\right) d\epsilon&=&-\left[\nu_{q}^{-1}\left(\tilde{\alpha},\epsilon\right)\exp\left(-\int_0^\epsilon \nu_{q}\left(\tilde{\alpha},s\right) ds\right)\right]_0^t \left[\tilde{Q}_{i,j}^{(3,3)}\right]_{\epsilon=0},
\end{eqnarray}
The transport coefficients of the IVHS, which are calculated by Chapman-Enskog method, are written as \cite{Yano}
\begin{eqnarray}
\mu\left(\alpha,\Omega\right)&=&\frac{p(\tilde{\alpha})}{\nu_\pi\left(\tilde{\alpha}\right) \left(1-\chi_p\right)},\\
\kappa\left(\alpha,\Omega\right)&=&\phi_T \frac{p\left(\tilde{\alpha}\right) R}{\nu_q\left(\tilde{\alpha}\right)(1-\chi_T)},\\
\eta\left(\alpha,\Omega\right)&=&\frac{1}{1-\chi_\rho}\left(\frac{\chi_T}{2(1-\chi_T)} \phi_T+\phi_\rho \right) \frac{p\left(\tilde{\alpha}\right)RT\left(\tilde{\alpha}\right)}{\rho\left(\tilde{\alpha}\right)}\frac{1}{\nu_q\left(\tilde{\alpha}\right)}.
\end{eqnarray}
where $\chi_p:=(\zeta/\nu_{\pi})(1-\Omega/2)$, $\chi_T:=2 \zeta/\nu_q$, $\chi_\rho:=(2-\Omega/2) \zeta/\nu_q$, $\phi_\rho:=a_4/6$ and $\phi_T:=5/2+a_4/3$.\\
$\mu\left(\alpha,\Omega\right)$, $\kappa\left(\alpha,\Omega\right)$ and $\eta\left(\alpha,\Omega\right)$ at $t$ cannot be expressed with $\int_0^t \tilde{Q}_{ij,lm}^{(2,2)}(\epsilon) d\epsilon$, $\left[\tilde{Q}_{ij,lm}^{(2,2)}\right]_{\epsilon=0}$, $\int_0^t \tilde{Q}_{i,j}^{(3,3)}(\epsilon) d\epsilon$ and $\left[\tilde{Q}_{i,j}^{(3,3)}\right]_{\epsilon=0}$, explicitly, because we cannot express $\nu_\pi$ with $\int_0^t \tilde{Q}_{ij,lm}^{(2,2)}(\epsilon) d\epsilon$ and $\left[\tilde{Q}_{ij,lm}^{(2,2)}\right]_{\epsilon=0}$ in Eq. (B11) and $\nu_q$ with $\int_0^t \tilde{Q}_{i,j}^{(3,3)}(\epsilon) d\epsilon$ and $\left[\tilde{Q}_{i,j}^{(3,3)}\right]_{\epsilon=0}$ in Eq. (B12). \textcolor{black}{Of course, $\mu\left(\alpha,1\right)$, $\kappa\left(\alpha,1\right)$ and $\eta\left(\alpha,1\right)$ in Eqs. (B13)-(B15) coincide with those calculated for the IHS by Brey \textit{et. al}. \cite{Brey4}.} On the other hand, we can obtain initial values of $\nu_\pi$ and $\nu_q$ by setting $t=\infty$ such as
\begin{eqnarray}
\nu_\pi\left(\tilde{\alpha},0\right)=\frac{\left[\tilde{Q}_{ij,lm}^{(2,2)}\right]_{\epsilon=0}}{\int_0^\infty \tilde{Q}_{ij,lm}^{(2,2)}\left(\epsilon\right) d\epsilon},~~\nu_q\left(\tilde{\alpha},0\right)=\frac{\left[\tilde{Q}_{i,j}^{(3,3)}\right]_{\epsilon=0}}{\int_0^\infty \tilde{Q}_{i,j}^{(3,3)}\left(\epsilon\right) d\epsilon}.
\end{eqnarray}
Substituting Eq. (B16) into Eqs. (B13)-(B15), we obtain the modified Green-Kubo expression for the transport coefficients using initial values of transport coefficients. Of course, such modified Green-Kubo expression obtained using initial values of transport coefficients are nonlinear responses. In the DSMC calculation, the transport coefficients at $t$ are obtained, when we calculate $\int_0^\infty \tilde{Q}_{ij,lm}^{(2,2)}(\epsilon) d\epsilon$ and $\int_0^\infty \tilde{Q}_{i,j}^{(3,3)}\left(\epsilon\right) d\epsilon$ by setting $t=0$.\\
Another approach to calculate transport coefficients for the granular gas under the HCS, we consider the map: $\epsilon \rightarrow \epsilon^\prime:=\left(\sqrt{RT_\infty}\left(T/T_\infty\right)^{\Omega/2}/L_\infty\right)^{-1}$, in which quantities with subscript $\infty$ correspond to representative values, in accordance with the map by Dufty and Brey \cite{Dufty}. This map nondimensionalizes dissipation rates $\nu_\pi$ and $\nu_q$, which depend on time, in time independent forms. Applying this map to Eqs. (B7) and (B8), we obtain $\int_0^\infty \tilde{Q}_{ij,lm}^{(2,2)}(\epsilon^\prime) d\epsilon^\prime$ and $\int_0^\infty \tilde{Q}_{i,j}^{(3,3)}\left(\epsilon^\prime\right) d\epsilon^\prime$ as
\begin{eqnarray}
\int_0^\infty \tilde{Q}_{ij,lm}^{(2,2)}\left(\epsilon^\prime\right) d\epsilon^\prime=\left[\tilde{Q}_{ij,lm}^{(2,2)}\right]_{\epsilon=0}\frac{1}{\tilde{\nu}_{\pi}\left(\tilde{\alpha}\right)},\\
\int_0^\infty \tilde{Q}_{i,j}^{(3,3)}\left(\epsilon^\prime\right) d\epsilon^\prime=\left[\tilde{Q}_{i,j}^{(3,3)}\right]_{\epsilon=0}\frac{1}{\tilde{\nu}_q\left(\tilde{\alpha}\right)},
\end{eqnarray}
where $\tilde{\nu}_\pi=\epsilon^\prime \nu_\pi$ and $\tilde{\nu}_q=\epsilon^\prime \nu_q$. As mentioned above, $\tilde{\nu}_\pi$ and $\tilde{\nu}_q$ are time independent. Substituting $\tilde{\nu}_\pi$ and $\tilde{\nu}_q$ in Eqs. (B17) and (B18) into Eqs. (B13)-(B15), we obtain another form of the modified Green-Kubo expression for the transport coefficients. In the DSMC calculation, we use the time interval $d \epsilon^\prime$ to integrate $\tilde{Q}_{ij,lm}^{(2,2)}\left(\epsilon^\prime\right)$ and $\tilde{Q}_{i,j}^{\left(3,3\right)}\left(\epsilon^\prime\right)$ in Eqs. (B17) and (B18), which are calculated using the temperature by each time step. The relation between the modified Green-Kubo expression for the transport coefficients, which was introduced on the basis of the two-point kinetic theory, and Green-Kubo expression by Dufty and Brey \cite{Brey} must be addressed in our future work.
\section{\textcolor{black}{Comments on other possibilities for kinetic calculation of transport coefficients}}
\textcolor{black}{The differences between $\kappa$ and $\eta$, which were obtained on the basis of Green-Kubo expression by Dufty and Brey \cite{Dufty}, and those obtained using the first order of approximation in Chapman-Enskog method, were numerically confirmed using the DSMC method by Brey \textit{et al} \cite{Brey}. Then, Garzo, Santos and Montanero \cite{Garzo2} proposed the modified Chapman-Enskog method to calculate the transport coefficients, $\mu$, $\kappa$ and $\eta$ together with the self-diffusion coefficient. In particular, $\kappa$ and $\eta$, which were calculated using the modified Chapman-Enskog method \cite{Garzo2} in a similar way to the method by Lutsko \cite{Lutsko}, are much more similar to $\kappa$ and $\eta$ \cite{Brey}, which were calculated on the basis of Green-Kubo expression by Dufty and Brey using the DSMC method, than $\kappa$ and $\eta$, which were calculated using the first order approximation in the conventional Chapman-Enskog method. Such modified Chapman-Enskog method expands $f\left(\bm{c}\right)$ around not $f_{\text{\tiny{MB}}}\left(\bm{c}\right)$ but $f^{(0)}\left(\bm{c}\right)$ in Eq. (2), where coefficients in the modified Sonine polynomial are determined to satisfy the orthogonality between two different Sonine polynomials. Provided that Grad's method is used to expand $f\left(\bm{c}\right)$ around $f^{(0)}\left(\bm{c}\right)$, the modified Chapman-Enskog method corresponds to the modified Grad's method, in which $f\left(\bm{c}\right)$ is expanded around $f^{(0)}\left(\bm{c}\right)$ in Eq. (2) using modified Hermite polynomials, whose coefficients in Hermite polynomials and coefficients on Hermite polynomials are determined using the orthogonality between two different modified Hermite polynomials or definitions of moments, respectively, such that coefficients $\varphi_{ij}^{(2)}$ and $\varphi_i^{(3)}$ in modified Hermite polynomials, $\hat{H}_{ij}^{(2)}=v_iv_j-\delta_{ij} \varphi_{ij}^{(2)}$ and $\hat{H}_{i}^{(3)}=v_i\left(v^2-5\varphi_i^{(3)}\right)$ are determined by their orthogonality, namely, $\int_{\mathbb{V}^3} \hat{H}_{ij}^{(2)} f^{(0)} d \bm{c}=0$ and $\int_{\mathbb{V}^3} \hat{H}_{i}^{(3)} H_i^{(1)} f^{(0)} d \bm{c}=0$ ($H_i^{(1)}=v_i$), and coefficients on Hermite polynomials $\varsigma_{ij}^{(2)}$ and $\varsigma_i^{(3)}$ are determined by the definition of moments, $p_{ij}^{(2)}/p=\rho^{-1}\int_{\mathbb{V}^3} H_{ij}^{(2)} f d \bm{c}$ and $q_i/(p\sqrt{RT})=1/2\rho^{-1}\int_{\mathbb{V}^3} H_i^{(3)} f d\bm{c}$. Such modified Hermite polynomials were also applied to the quantum gas by the author \cite{Yano2}. We, however, remind that such modified Sonine or Hermite polynomials do not always satisfy the completeness of the expansion of $f\left(\bm{c}\right)$ in a mathematical sense. For instance, $f\left(\bm{c}\right)$ is expanded around $f^{(0)}\left(\bm{c}\right)$ in Eq. (2) using modified Hermite polynomials, namely, $\hat{H}_\nu^{(n)}$, such as
\begin{eqnarray}
&&f\left(\bm{c}\right) \simeq f_{\text{\tiny{MB}}}\left(\bm{c}\right)\left(1+\frac{\varsigma_{ij}^{(2)}}{2}\frac{p_{ij}}{p}\hat{H}_{ij}^{(2)}+\frac{\varsigma_i^{(3)} q_i \hat{H}_i^{(3)}}{5p\sqrt{RT}}\right)\left(1+\frac{1}{120}a_4 H^{(4)}+\frac{1}{5400}a_6 H^{(6)}\right), \nonumber \\
&&\text{where}~~~ \varsigma_{ij}^{(2)}=1,~~\hat{H}_{ij}^{(2)}=H_{ij}^{(2)},~~\varsigma_i^{(3)}=\frac{90}{90+33a_4-a_4^2+3a_6},~~\varphi_i^{(3)}=1+\frac{a_4}{15}.
\end{eqnarray}
From Eq. (C1), we can calculate $\mu\left(\alpha,\Omega\right)$, $\kappa\left(\alpha,\Omega\right)$ and $\eta\left(\alpha,\Omega\right)$ for the IVHS, whereas such calculations of the transport coefficients will be described elsewhere.\\
As one possibility for the improvement of differences between $\kappa\left(\alpha,\Omega\right)$ and $\eta\left(\alpha,\Omega\right)$, which are calculated by Eqs. (38)-(40), and $\kappa\left(\alpha,\Omega\right)$ and $\eta\left(\alpha,\Omega\right)$, which are calculated by Eqs. (41)-(43) using the DSMC method, nonequilibrium moments are considered in the definition of the cooling rate. In this paper, we used the cooling rate $\zeta$ in Eq. (4), which never depends on nonequilibrium moments. Provided that $f\left(\bm{c}\right)=f^{(0)}\left(\bm{c}\right)$ in Eq. (2), the cooling rate is obtained by neglecting all the nonlinear terms as
\begin{eqnarray}
\zeta^\prime\left(\alpha,\Omega\right)=\frac{5}{2\left(5+\Omega\right)\tau}\left(1-\alpha^2\right)\left[1+\underbrace{\frac{\Omega\left(2+\Omega\right)}{240} a_4-\frac{\Omega\left(4-\Omega^2\right)}{2160}a_6}_{\hat{\zeta}_{\text{\tiny{neq}}}} \right],
\end{eqnarray}
where $\zeta^{\prime}\left(\alpha,1\right)$ for the IHS is same as the cooling rate for the IHS, which was calculated by Brilliantov and P$\ddot{\mbox{o}}$schel \cite{Brilliantov}.\\
Here, we must confirm $\hat{\zeta}_{\text{\tiny{neq}}} \ll 1$ to validate $\zeta$ in Eq. (4), which was used for the IHS, IVHS with $\Omega=0.6$ and IMS in our analytical results. Figure 11 shows $\hat{\zeta}_{\text{\tiny{neq}}}$ versus $\alpha$ in cases of the IHS, IVHS with $\Omega=0.6$ and IMS, when $a_4$ and $a_6$ in Eq. (C1) are calculated by Eqs. (6) and (7). As shown in Fig. 11, $\hat{\zeta}_{\text{\tiny{neq}}}\ll 1$ is surely obtained in cases of the IHS, IVHS with $\Omega=0.6$ and IMS. In particular, $\zeta$ does not depend on nonequilibrium moments in the case of the IMS. Therefore, the modification of $\zeta$ in Eq. (4) with $\zeta^\prime$ in Eq. (C1) does not improve differences between $\kappa\left(\alpha,\Omega\right)$ and $\eta\left(\alpha,\Omega\right)$, which are calculated by Eqs. (38)-(40), and $\kappa\left(\alpha,\Omega\right)$ and $\eta\left(\alpha,\Omega\right)$, which are calculated by Eqs. (41)-(43) using the DSMC method.}\\
\textcolor{black}{As other possibility for the improvement of differences between $\kappa\left(\alpha,\Omega\right)$ and $\eta\left(\alpha,\Omega\right)$, which are calculated by Eqs. (38)-(40), and $\kappa\left(\alpha,\Omega\right)$ and $\eta\left(\alpha,\Omega\right)$, which are calculated by Eqs. (41)-(43) using the DSMC method, nonlinear collisional moments are considered.}\\ 
\textcolor{black}{For example, moment equations of $p_{ij}$ and $q_i$ for the IVHS under Gaussian thermostat are written, when we substitute Grad's 14 moment equation, namely, $f\left(\bm{c}\right)=f_{14}\left(\bm{c}\right)=f_{\text{\tiny{MB}}}\left(\bm{c}\right)\left[1+p_{ij}/(2p)H_{ij}^{(2)}+q_i H_i^{(3)}/\left(5p\sqrt{RT}\right)+a_4 H^{(4)}/120\right]$ into Eq. (1), multiply $C_i C_j-\delta_{ij}C^2/3$ and $C_i C^2/2$ by both sides of Eq. (1) and integrate over $\mathbb{V}^3$, as \cite{Yano}}
\begin{eqnarray}
&& \textcolor{black}{\frac{\partial p_{ij}}{\partial t}+\frac{\partial p_{ij}u_k}{\partial x_k}+\frac{4}{5}\frac{\partial q_{<i}}{\partial x_{j>}}+2p\frac{\partial u_{<i}}{\partial x_{j>}}+2p_{k<i}\frac{\partial u_{j>}}{\partial x_k}=\left[\zeta-\nu_{\pi}\left(1+\underbrace{\frac{1}{480}\left(\Omega-2\right)\Omega a_4}_{\beta_p}\right)\right] p_{ij},}\nonumber \\\\
&&\textcolor{black}{\frac{\partial q_i}{\partial t}+\frac{\partial q_i u_k}{\partial x_k}+\frac{5}{2}p\frac{\partial RT}{\partial x_i}+\frac{5}{2}p_{ik}\frac{\partial RT}{\partial x_k}+RT\frac{\partial p_{ik}}{\partial x_k}-RTp_{ik}\frac{\partial \ln \rho}{\partial x_k}} \nonumber \\
&&\textcolor{black}{-\frac{p_{ij}}{\rho}\frac{\partial p_{jk}}{\partial x_k}+\frac{7}{5}q_k\frac{\partial u_i}{\partial x_k}+\frac{2}{5}q_k\frac{\partial u_k}{\partial x_i}+\frac{2}{5}q_i\frac{\partial u_k}{\partial x_k}+\frac{1}{6}\frac{\partial s}{\partial x_i}} \nonumber \\
&&\textcolor{black}{=\left[\frac{3}{2}\zeta-\nu_q\left(1+\underbrace{\frac{\left(\Omega-2\right)\Omega\left(50+\alpha(\Omega-10)+7 \Omega\right)}{480\left(\alpha\left(70+29\Omega\right)-110-37\Omega\right)}a_4}_{\beta_q}\right) \right]q_i,}
\end{eqnarray}
\textcolor{black}{where $A_{\left<ij\right>}$ is the traceless tensor and $s:=\rho (RT)^2 a_4$.}\\
\textcolor{black}{Effects of nonlinear collisional moments are markedly small owing to $\left|\beta_p\right| \ll 1$ and $\left|\beta_q\right| \ll 1$ in Eqs. (C3) and (C4), as described in the author's previous study \cite{Yano}. In particular, $\beta_p=\beta_q=0$ is always obtained for the IMS, because of $\Omega=0$ in Eqs. (C3) and (C4). From one to one correspondence between $p_{ij}$ in Eq. (28) and $Q_{ij,lm}^{(2,2)}$ in Eq. (C3) or $q_i$ in Eq. (29) and $Q_{i,j}^{(3)}$ in Eq. (C3), inclusions of nonlinear collisional moments in Eqs. (28) and (29) do not improve differences between $\kappa(\alpha,\Omega)$ and $\eta(\alpha,\Omega)$, which are calculated by Eqs. (38)-(40), and $\kappa\left(\alpha,\Omega\right)$ and $\eta\left(\alpha,\Omega\right)$, which are calculated by Eqs. (41)-(43) using the DSMC method.}\\
\textcolor{black}{From above discussions, the only remained way to improve differences between $\kappa\left(\alpha,\Omega\right)$ and $\eta\left(\alpha,\Omega\right)$, which are calculated by Eqs. (38)-(40), and $\kappa\left(\alpha,\Omega\right)$ and $\eta\left(\alpha,\Omega\right)$, which are calculated by Eqs. (41)-(43) using the DSMC method, is to expand $f\left(\bm{c}\right)$ around $f^{(0)}\left(\bm{c}\right)$ in Eq. (2) using the modified Sonine polynomials or modified Hermite polynomials such as Eq. (C1), when we restrict ourselves to the form of $f^{(0)}\left(\bm{c}\right)$ to $f^{(0)}\left(\bm{c}\right)$ in Eq. (2). Therefore, the expansion of $f\left(\bm{c}\right)$ around $f^{(0)}\left(\bm{c}\right)$ is worthy of trying to improve differences between $\kappa(\alpha,\Omega)$ and $\eta(\alpha,\Omega)$, which are calculated by Eqs. (38)-(40), and $\kappa(\alpha,\Omega)$ and $\eta(\alpha,\Omega)$, which are calculated by Eqs. (41)-(43) using the DSMC method. As described above, such an expansion of $f\left(\bm{c}\right)$ around $f^{(0)}\left(\bm{c}\right)$ is, however, artificial from the viewpoint of the completeness of the expansion, whereas all the modified Sonine or Hermite polynomials are presumably insufficient basis to cover the functional space mapped by $f\left(\bm{c}\right) (\bm{c}\in \mathbb{V}^3)$, because the expansion of $f\left(\bm{c}\right)$ around $f_{\text{\tiny{MB}}}\left(\bm{c}\right)$ with Sonine or Hermite polynomials is complete for the elastic gas owing to Gaussian form of $f_{\text{\tiny{MB}}}\left(\bm{c}\right)$, accidentally.}
\end{appendix}

\newpage
\hspace{-1em}\textsf{Figure captions}:\\
FIG. 1: $a_4$ versus $\alpha$ for the IHS (left frame). $a_6$ versus $\alpha$ for the IHS (right frame).\\
FIG. 2: $a_4$ versus $\alpha$ for the IVHS with $\Omega=0.6$ (left frame). $a_6$ versus $\alpha$ for the IVHS with $\Omega=0.6$ (right frame).\\
FIG. 3: $a_4$ versus $\alpha$ for the IMS (left frame). $a_6$ versus $\alpha$ for the IMS (right frame).\\
FIG. 4: $\psi_2\left(\alpha\right)$ versus $\alpha$ (left frame) and $\psi_3\left(\alpha\right)$ versus $\alpha$ (right frame) for the IHS.\\
FIG. 5: $\psi_2\left(\alpha\right)$ versus $\alpha$ (left frame) and $\psi_3\left(\alpha\right)$ versus $\alpha$ (right frame) for the IVHS with $\Omega=0.6$.\\
FIG. 6: $\psi_2\left(\alpha\right)$ versus $\alpha$ (left frame) and $\psi_3\left(\alpha\right)$ versus $\alpha$ (right frame) for the IMS.\\
FIG. 7: $\phi_2\left(\hat{\epsilon}\right)$ and $\phi_3\left(\hat{\epsilon}\right)$ versus $\hat{\epsilon}$ for the IHS.\\
FIG. 8: $\phi_2\left(\hat{\epsilon}\right)$ and $\phi_3\left(\hat{\epsilon}\right)$ versus $\hat{\epsilon}$ for the IVHS with $\Omega=0.6$.\\
FIG. 9: $\phi_2\left(\hat{\epsilon}\right)$ and $\phi_3\left(\hat{\epsilon}\right)$ versus $\hat{\epsilon}$ for the IMS.\\
FIG. 10: $\mu\left(\alpha\right)/\mu(0)$, $\kappa\left(\alpha\right)/\kappa(0)$ and $\tilde{\eta}\left(\alpha\right)$ versus $\alpha$ for the IHS, IVHS with $\Omega=0.6$ and IMS.\\
FIG. 11: $\hat{\zeta}_{\text{\tiny{neq}}}$ versus $\alpha$ in cases of the IHS, IVHS with $\Omega=0.6$ and IMS.
\newpage
\begin{center}
\includegraphics[width=1.0\textwidth]{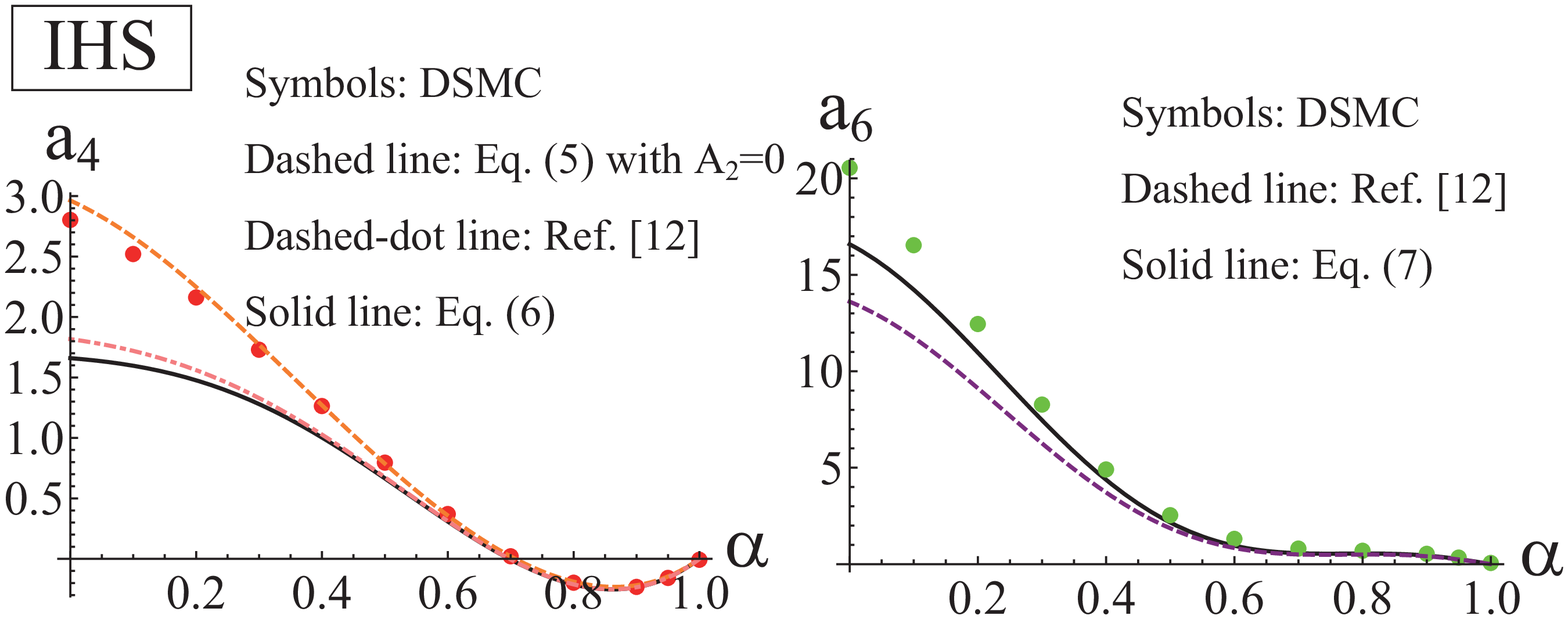}\\
\footnotesize{FIG. 1: $a_4$ versus $\alpha$ for the IHS (left frame). $a_6$ versus $\alpha$ for the IHS (right frame).}
\end{center}
\newpage
\begin{center}
\includegraphics[width=1.0\textwidth]{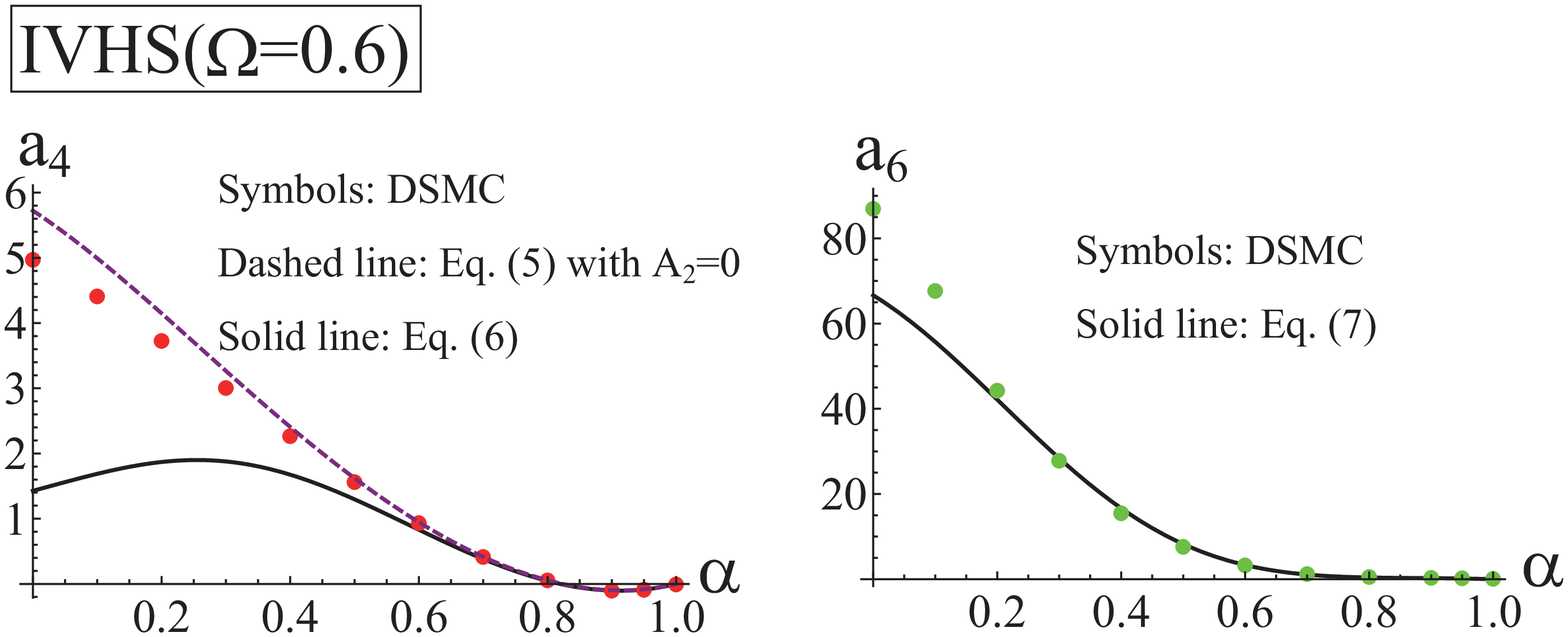}\\
\footnotesize{FIG. 2: $a_4$ versus $\alpha$ for the IVHS with $\Omega=0.6$ (left frame). $a_6$ versus $\alpha$ for the IVHS with $\Omega=0.6$ (right frame).}
\end{center}
\newpage
\begin{center}
\includegraphics[width=1.0\textwidth]{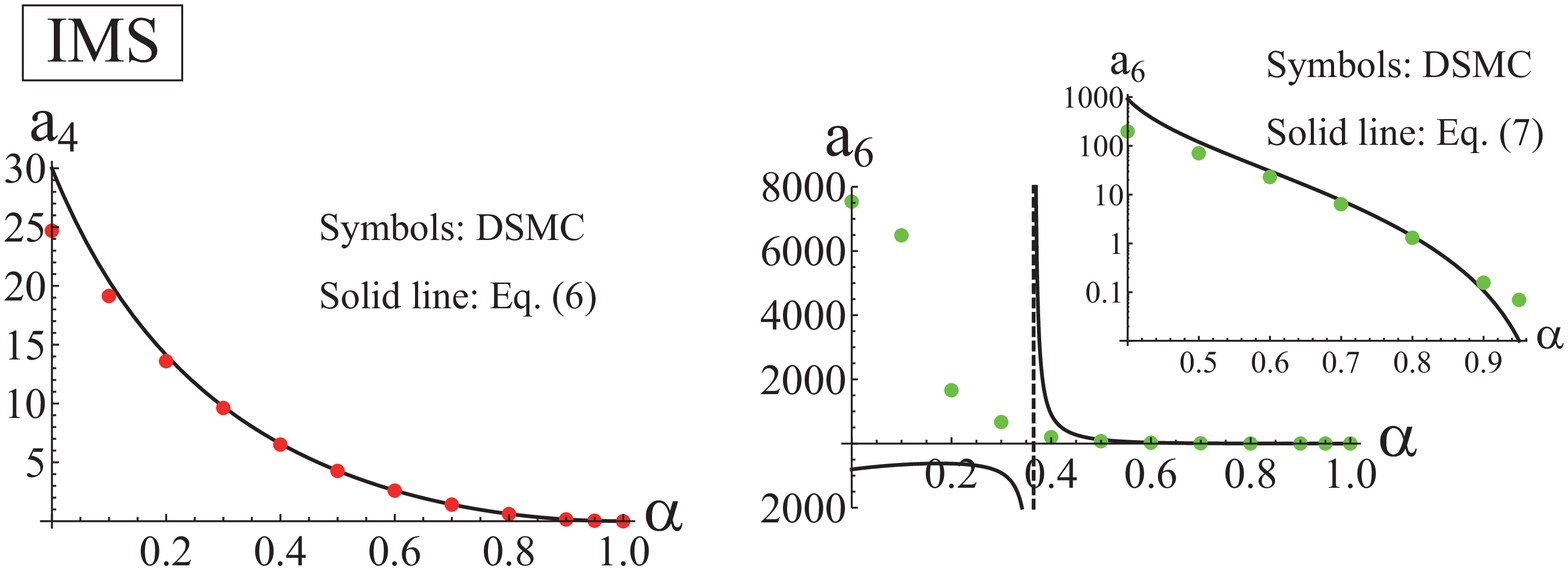}\\
\footnotesize{FIG. 3: $a_4$ versus $\alpha$ for the IMS (left frame). $a_6$ versus $\alpha$ for the IMS (right frame).}
\end{center}
\newpage
\begin{center}
\includegraphics[width=1.0\textwidth]{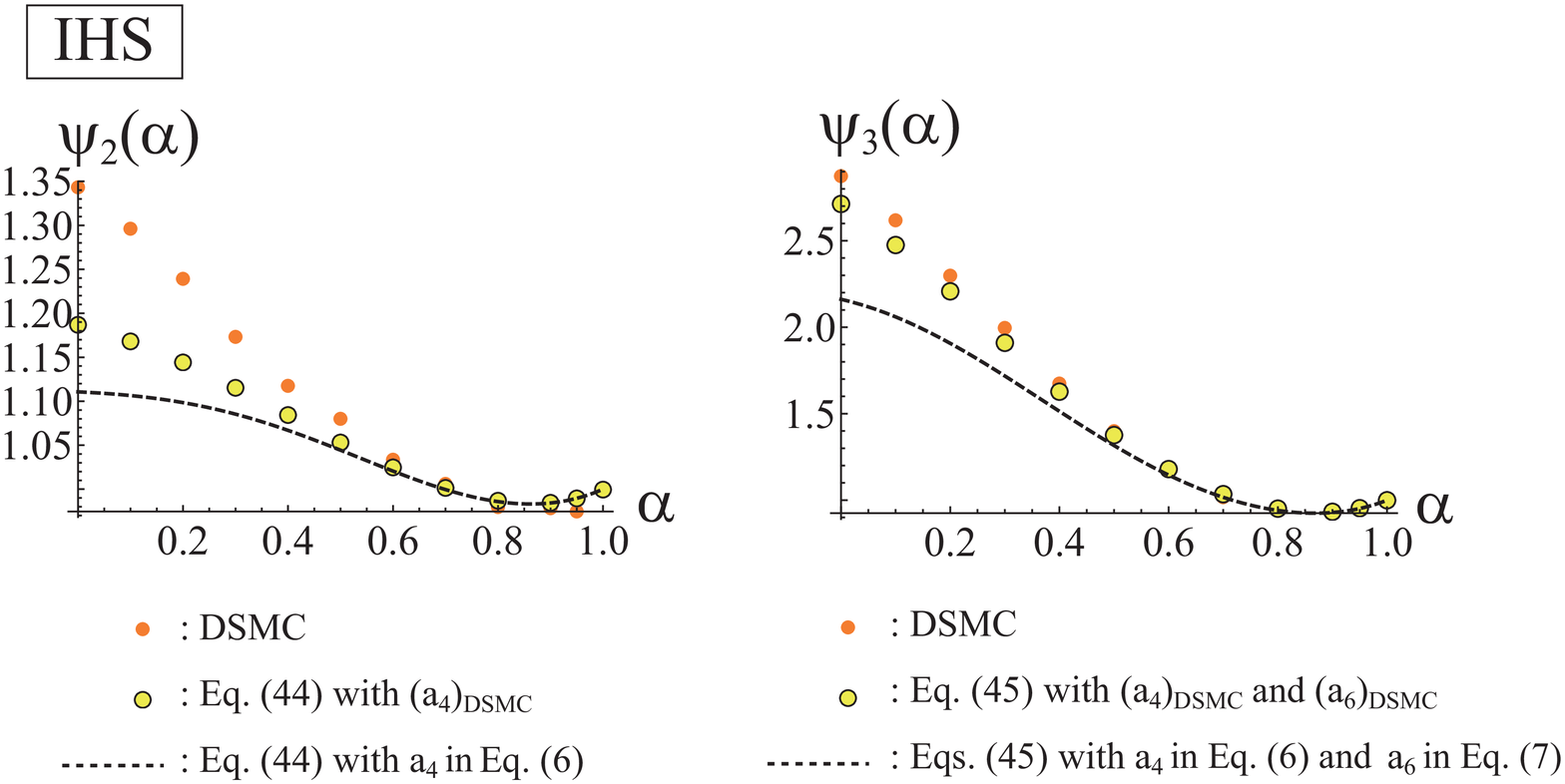}\\
\footnotesize{FIG. 4: $\psi_2\left(\alpha\right)$ versus $\alpha$ (left frame) and $\psi_3\left(\alpha\right)$ versus $\alpha$ (right frame) for the IHS.}
\end{center}
\newpage
\begin{center}
\includegraphics[width=1.0\textwidth]{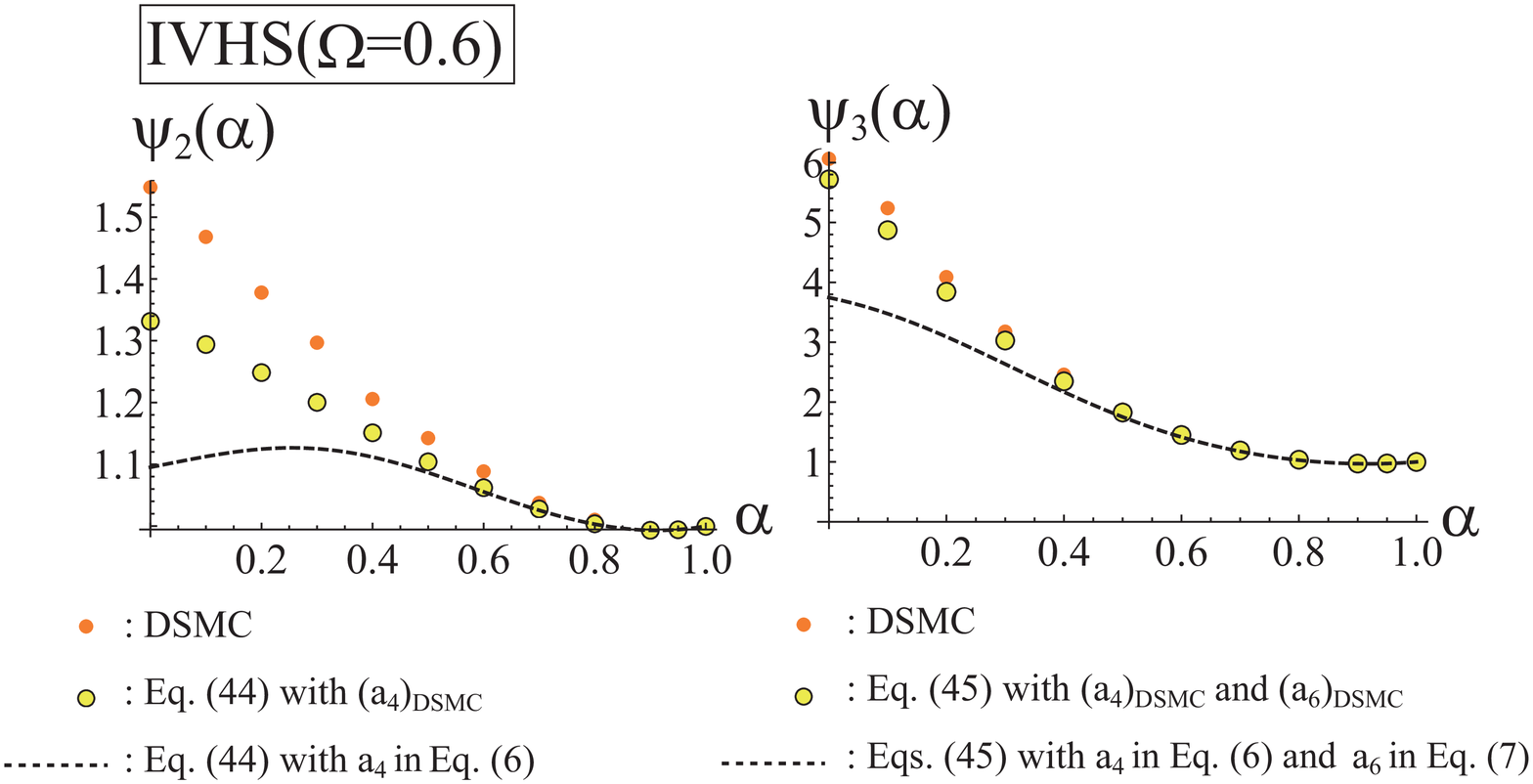}\\
\footnotesize{FIG. 5: $\psi_2\left(\alpha\right)$ versus $\alpha$ (left frame) and $\psi_3\left(\alpha\right)$ versus $\alpha$ (right frame) for the IVHS with $\Omega=0.6$.}
\end{center}\newpage
\begin{center}
\includegraphics[width=1.0\textwidth]{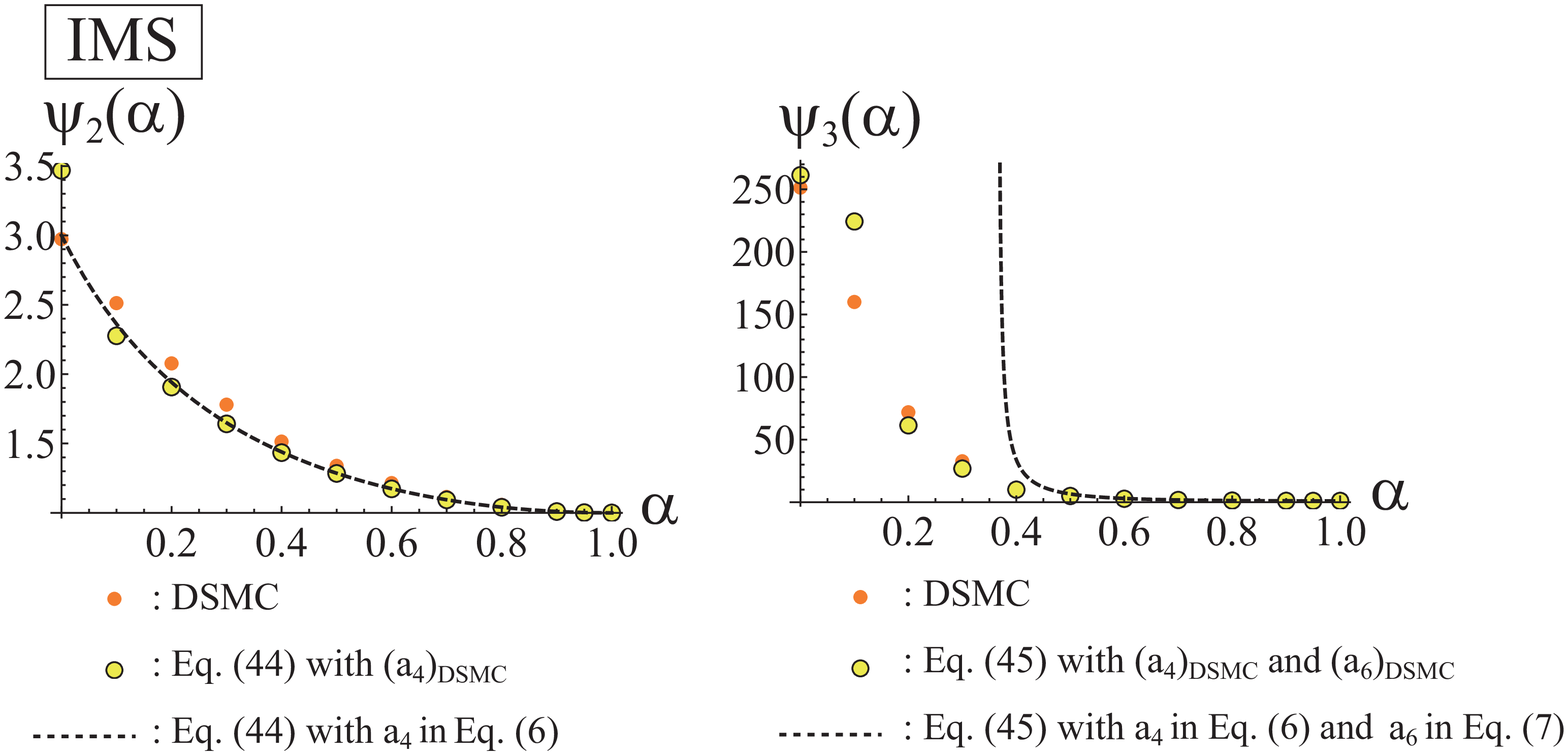}\\
\footnotesize{FIG. 6: $\psi_2\left(\alpha\right)$ versus $\alpha$ (left frame) and $\psi_3\left(\alpha\right)$ versus $\alpha$ (right frame) for the IMS.}
\end{center}
\newpage
\begin{center}
\includegraphics[width=1.0\textwidth]{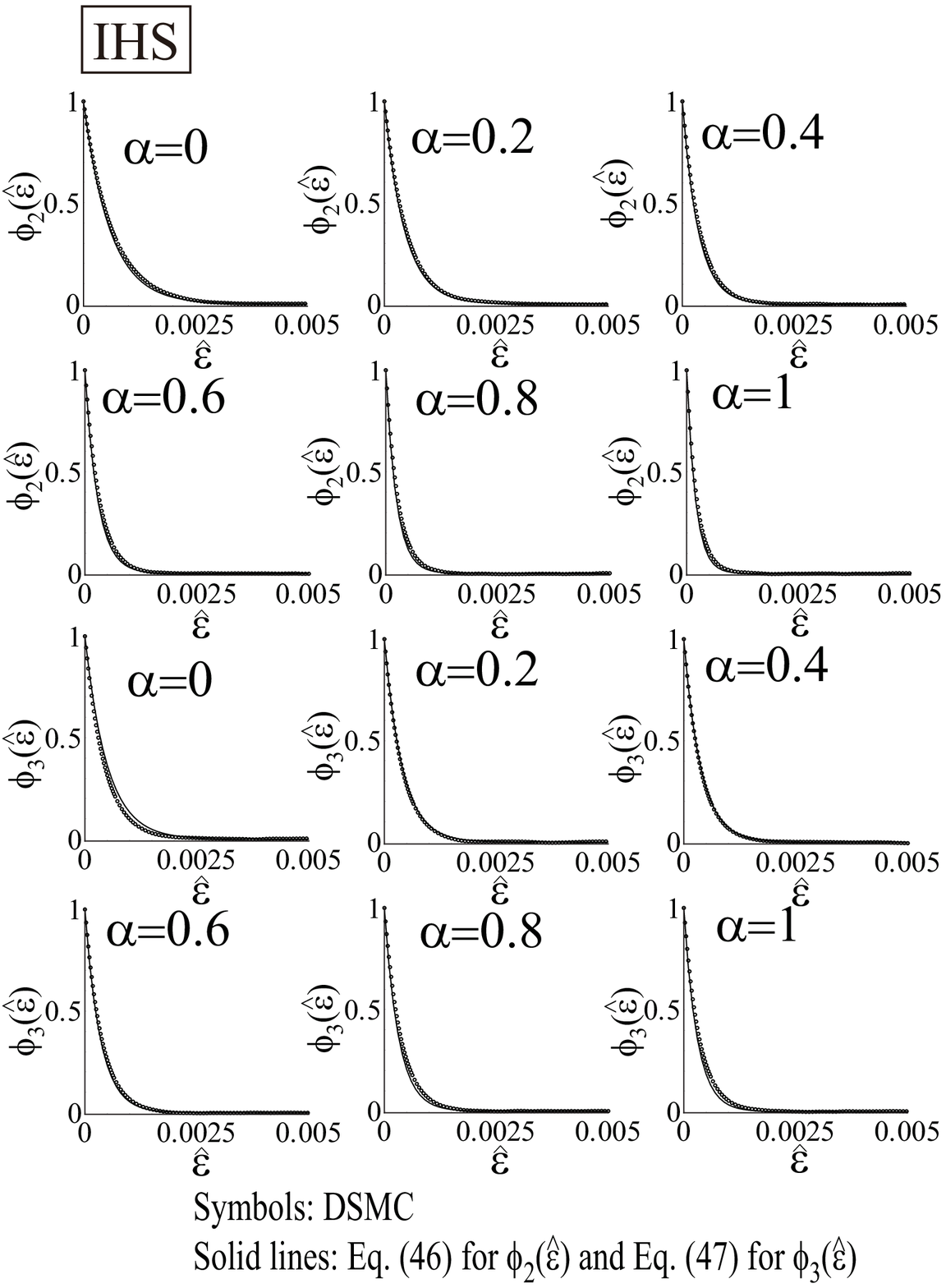}\\
\footnotesize{FIG. 7: $\phi_2(\hat{\epsilon})$ and $\phi_3(\hat{\epsilon})$ versus $\hat{\epsilon}$ for the IHS.}
\end{center}
\newpage
\begin{center}
\includegraphics[width=1.0\textwidth]{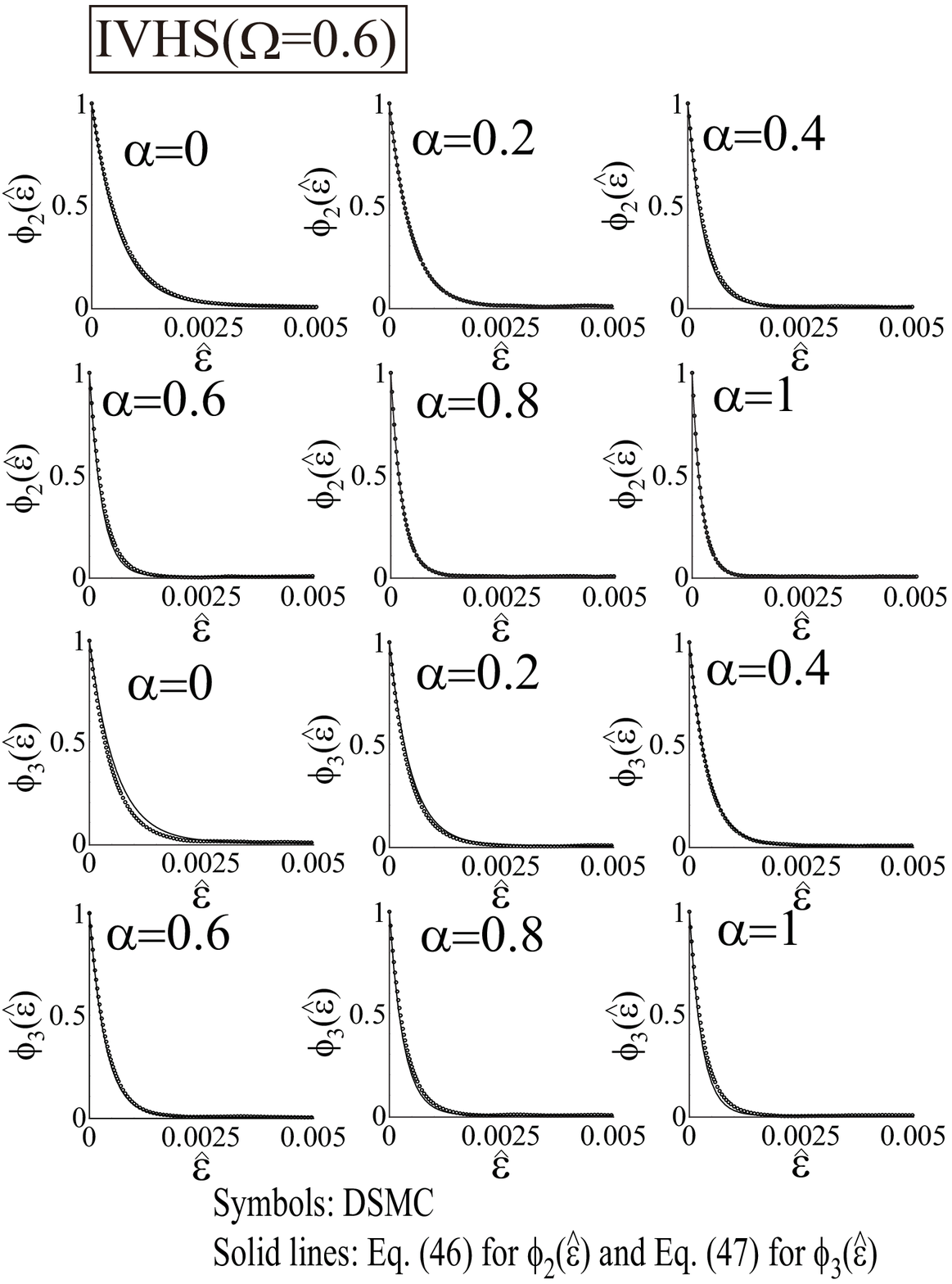}\\
\footnotesize{FIG. 8: $\phi_2(\hat{\epsilon})$ and $\phi_3(\hat{\epsilon})$ versus $\hat{\epsilon}$ for the IVHS with $\Omega=0.6$.}
\end{center}
\newpage
\begin{center}
\includegraphics[width=1.0\textwidth]{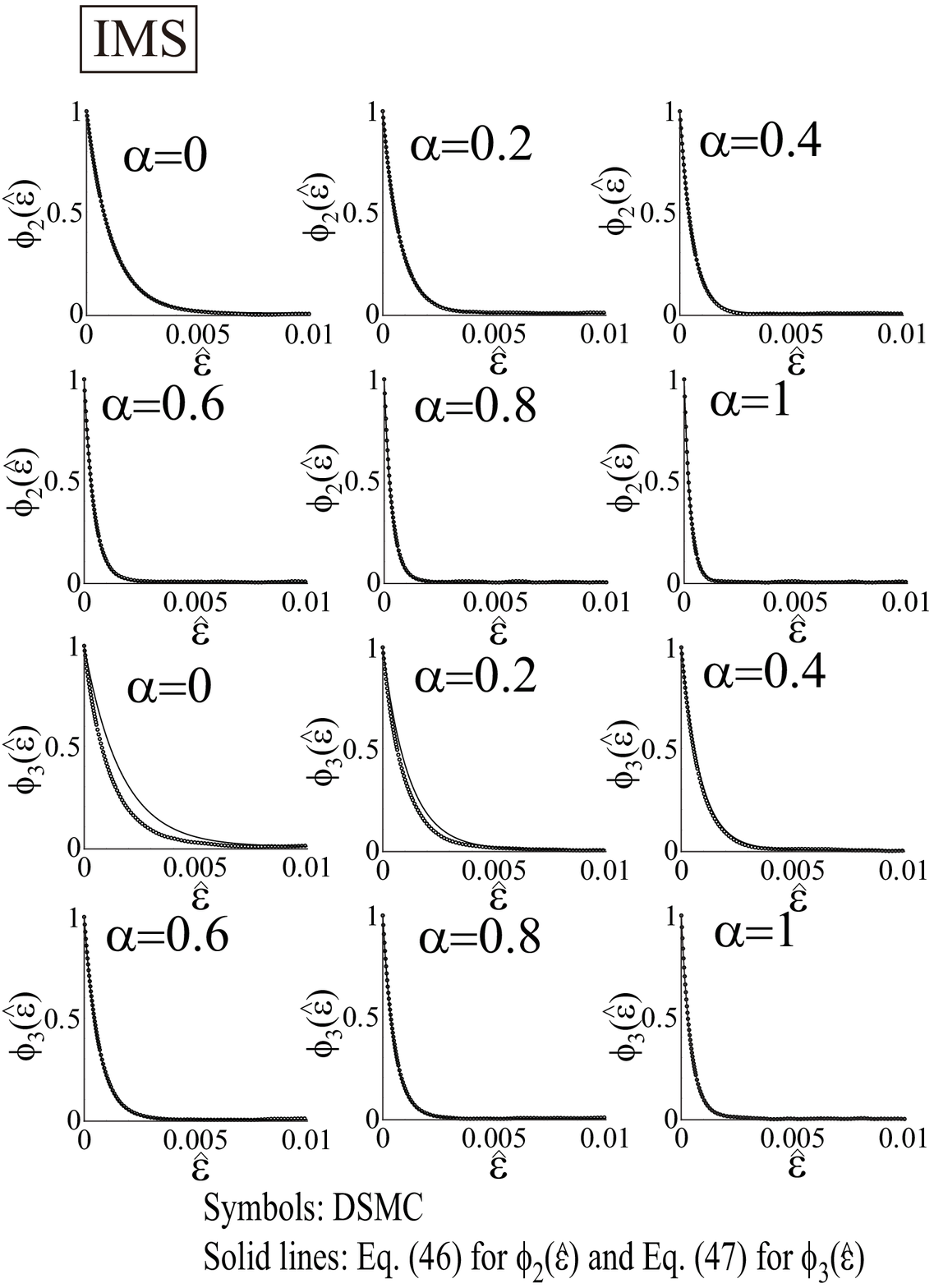}\\
\footnotesize{FIG. 9: $\phi_2(\hat{\epsilon})$ and $\phi_3(\hat{\epsilon})$ versus $\hat{\epsilon}$ for the IMS.}
\end{center}
\newpage
\begin{center}
\includegraphics[width=1.0\textwidth]{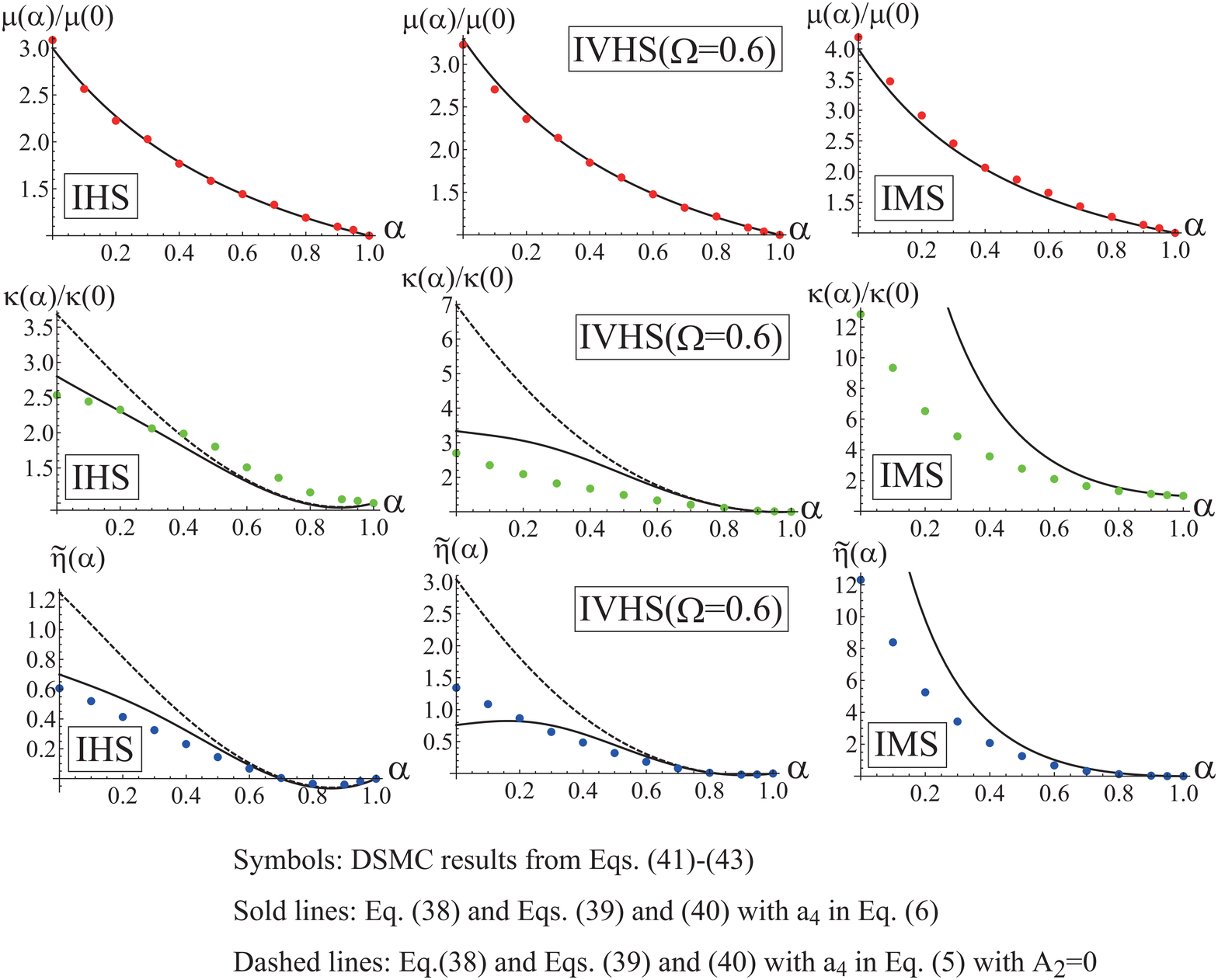}\\
\footnotesize{FIG. 10: $\mu(\alpha)/\mu(0)$, $\kappa(\alpha)/\kappa(0)$ and $\tilde{\eta}(\alpha)$ versus $\alpha$ for the IHS, IVHS with $\Omega=0.6$ and IMS.}
\end{center}
\newpage
\begin{center}
\includegraphics[width=0.5\linewidth]{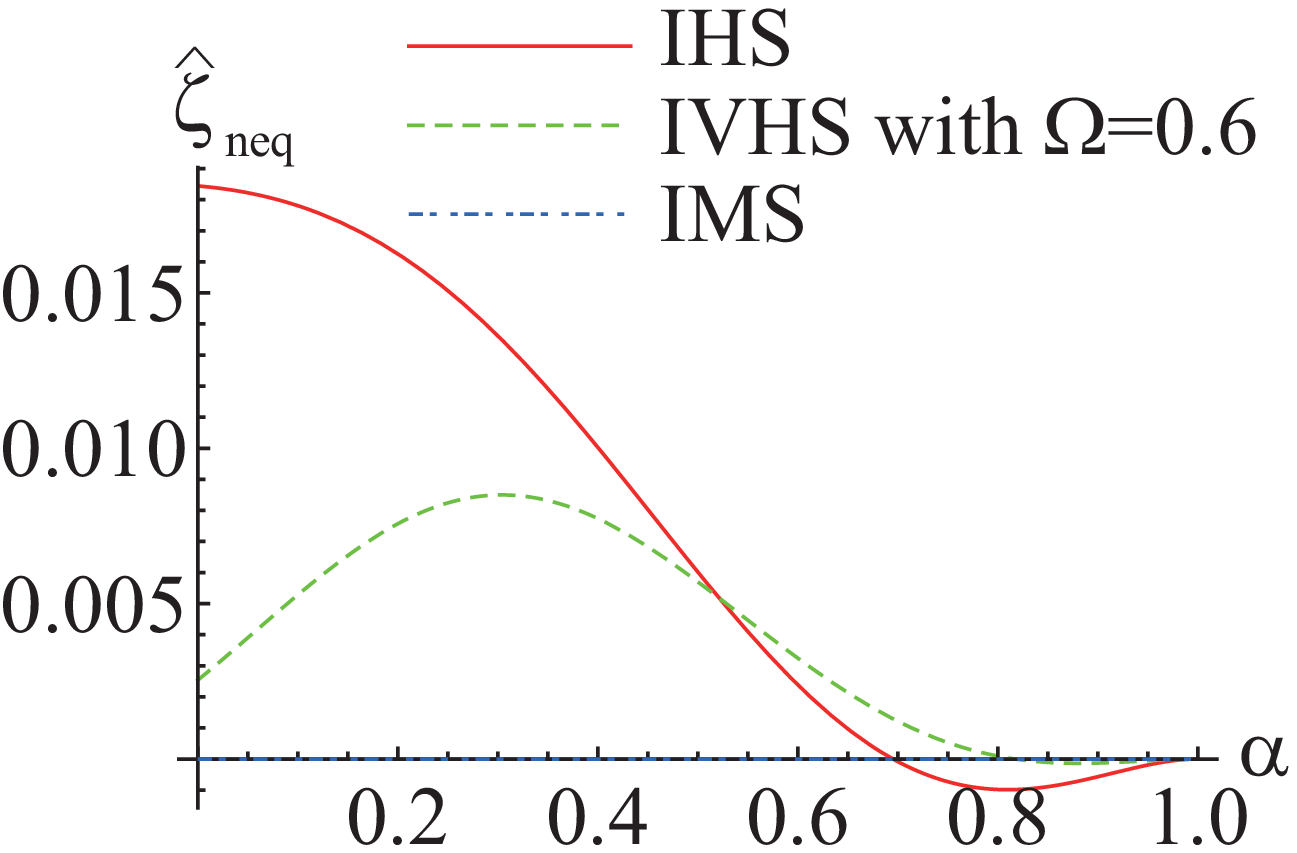}\\
\footnotesize{FIG. 11:  $\hat{\zeta}_{\text{\tiny{neq}}}$ versus $\alpha$ in cases of the IHS, IVHS with $\Omega=0.6$ and IMS.}
\end{center}

\begin{thebibliography}{23}
\bibitem{memo} \textcolor{black}{Readers remind that the time evolution of $a_6$ in Eq. (5) is not a self-similar form unlike the time evolution of $a_4$ in the case of the IMS, because $A_2=0 $ and $B_1 \neq 0$ are obtained in the case of the IMS, as shown in Eqs. (A3) and (A5). Thus, we conjecture nonlinear terms with $a_4$ remain as nonzero terms in the case of the IMS.}
\end{thebibliography}
\end{document}